\documentclass[12pt,ams]{article}
\usepackage{graphicx}
\usepackage{amsmath}

\begin{document}

\date{}
\title{\textbf{A study of the maximal Abelian gauge in $SU(2)$ Euclidean Yang-Mills
theory in the presence of the Gribov horizon} }
\author{\textbf{M.A.L. Capri}\thanks{
marcio@dft.if.uerj.br} \ , \textbf{V.E.R. Lemes}\thanks{
vitor@dft.if.uerj.br} \ , \textbf{R.F. Sobreiro}\thanks{
sobreiro@uerj.br}{\ }\ , \and  \textbf{S.P. Sorella}\thanks{
sorella@uerj.br}{\ }\footnote{Work supported by FAPERJ, Funda{\c
c}{\~a}o de Amparo {\`a} Pesquisa do Estado do Rio de Janeiro,
under the program {\it Cientista do Nosso Estado},
E-26/151.947/2004.}  \ , \textbf{R. Thibes}\thanks{
thibes@dft.if.uerj.br} \\
\\
\textit{UERJ, Universidade do Estado do Rio de Janeiro,} \and \textit{Rua S{%
\~a}o Francisco Xavier 524, 20550-013 Maracan{\~a},} \and \textit{Rio de
Janeiro, Brasil}}
\maketitle

\begin{abstract}
We pursue the study of $SU(2)$ Euclidean Yang-Mills theory in the
maximal Abelian gauge by taking into account the effects of the
Gribov horizon. The Gribov approximation, previously introduced in
\cite{Capri:2005tj}, is improved through the introduction of the
horizon function, which is constructed under the requirements of
localizability and renormalizability. By following Zwanziger's
treatment of the horizon function in the Landau gauge, we prove
that, when cast in local form, the horizon term of the maximal
Abelian gauge leads to a quantized theory which enjoys
multiplicative renormalizability, a feature which is established
to all orders by means of the algebraic renormalization.
Furthermore, it turns out that the horizon term is compatible with
the local residual $U(1)$ Ward identity, typical of the maximal
Abelian gauge, which is easily derived. As a consequence, the
nonrenormalization theorem, $Z_{g}Z_{A}^{1/2}=1$, relating the
renormalization factors of the gauge coupling constant $Z_{g}$ and
of the diagonal gluon field $Z_{A}$, still holds in the presence
of the Gribov horizon. Finally, we notice that a generalized
dimension two gluon operator can be also introduced. It is $BRST$\
invariant on-shell, a property which ensures its multiplicative
renormalizability. Its anomalous dimension is not an independent
parameter of the theory, being obtained from the renormalization
factors of the gauge coupling constant and of the diagonal
antighost field.
\end{abstract}

\newpage

\section{Introduction}

The maximal Abelian gauge \cite
{'tHooft:1981ht,Kronfeld:1987vd,Kronfeld:1987ri} provides a useful tool in
order to investigate nonperturbative aspects of Yang-Mills theories. This
gauge turns out to be suitable for the study of the dual superconductivity
mechanism for color confinement \cite{scon}, according to which Yang-Mills
theories in the low energy region should be described by an effective
Abelian theory \cite{Ezawa:bf,Suzuki:1989gp,Suzuki:1992gz,Hioki:1991ai} in
the presence of monopoles. The condensation of these magnetic charges leads
to a dual Meissner effect resulting in quark confinement. Here, the Abelian
configurations are identified with the diagonal components $A_{\mu }^{i}$, $%
i=1,...,N-1$, of the gauge field corresponding to the $\left(
N-1\right) $ generators of the Cartan subgroup of $SU(N)$. The
remaining off-diagonal components $A_{\mu }^{a}$,
$a=1,...,N^{2}-N$, corresponding to the $\left( N^{2}-N\right) $
off-diagonal generators of $SU(N)$, are expected to acquire a mass
through a dynamical mechanism, thus decoupling at low energies.
\newline
\newline
The maximal Abelian gauge is a renormalizable gauge in the continuum \cite
{Min:1985bx,Fazio:2001rm,Dudal:2004rx,Gracey:2005vu}, while possessing a
lattice formulation \cite{Kronfeld:1987vd,Kronfeld:1987ri}. This property,
shared only by few other gauges, such as the Landau and Coulomb\footnote{%
In the case of the Coulomb gauge we should remind that the issue of the
renormalizability is still under debate, see \cite{Baulieu:1998kx} for a
detailed account.} gauges, gives to the maximal Abelian gauge a privileged
role. It allows for a comparison between the results obtained in the
continuum and in lattice numerical simulations. \newline
\newline
Numerical studies of the behavior of the gluon propagator in the case of $%
SU(2)$ can be found in \cite{Amemiya:1998jz,Bornyakov:2003ee}. In
particular, according to \cite{Bornyakov:2003ee}, both diagonal
and off-diagonal gluon propagators turn out to be suppressed in
the infrared region. The diagonal propagator exhibits a Gribov
like behavior, while the off-diagonal transverse component is of
the Yukawa type. The value reported for the off-diagonal gluon
mass is of approximately $1.2GeV$
\cite{Amemiya:1998jz,Bornyakov:2003ee}, and turns out to be twice
bigger than the mass parameter entering the diagonal propagator.
As a consequence, the off-diagonal propagator is short-ranged as
compared to the diagonal one, a feature which is in agreement with
the Abelian dominance hypothesis \cite
{Ezawa:bf,Suzuki:1989gp,Suzuki:1992gz,Hioki:1991ai}. \newline
\newline
As the Landau and Coulomb gauges, the maximal Abelian gauge is affected by
the existence of the Gribov copies \cite{Gribov:1977wm}\footnote{%
See \cite{Bruckmann:2000xd} for the construction of an explicit example of a
zero mode of the Faddeev-Popov operator.}, which have to be taken into
account in order to properly quantize the theory. A first step in this
direction was achieved in \cite{Capri:2005tj}, where the original
construction outlined by Gribov for the Landau gauge \cite{Gribov:1977wm}
was generalized to the maximal Abelian gauge. In particular, we have been
able to generalize to the maximal Abelian gauge Gribov's result stating that
for any field close to a horizon there is a gauge copy, close to the same
horizon, located on the other side of the horizon \cite{Gribov:1977wm}. This
result has provided support for restricting the domain of integration in the
Feynman path integral to the so-called Gribov region, \textit{i.e.} to the
region in field space whose boundary is the first Gribov horizon, where the
first vanishing eigenvalue of the Faddeev-Popov operator appears. \newline
\newline
According to Gribov's original procedure \cite{Gribov:1977wm}, this
restriction can be implemented by means of a no-pole condition on the ghost
two-point function, which can be worked out order by order \cite
{Gribov:1977wm}. In particular, the so called Gribov approximation amounts
to work out the no-pole condition at the first nontrivial order, resulting
in a new nonlocal term which has to be added to the Yang-Mills action. In
the Landau gauge, this term reads \cite{Gribov:1977wm}
\begin{equation}
S_{G}=-g^{2}\gamma ^{4}N\int d^{4}x\;A_{\mu }^{A}\frac{1}{\partial ^{2}}%
A_{\mu }^{A}\;,  \label{i1}
\end{equation}
where the index $A$ belongs to the adjoint representation of $SU(N)$, $%
A=1,....,N^{2}-1$, and where $\gamma $ is the Gribov parameter \cite
{Gribov:1977wm}. It is not a free parameter of the theory, being defined by
a gap equation which enables us to express it in terms of the gauge coupling
$g$. At the first order, the gap equation defining $\gamma $ is given by
\begin{equation}
1=\frac{3}{4}Ng^{2}\int \frac{d^{4}k}{\left( 2\pi \right) ^{4}}\frac{1}{%
k^{4}+2Ng^{2}\gamma ^{4}}\;.  \label{i2}
\end{equation}
Although expression $\left( \ref{i1}\right) $, being quadratic in the gauge
fields $A_{\mu }^{A}$, is useful to analyse the modifications of the tree
level gluon propagator due to the restriction to the Gribov region, it
suffers from several limitations. For instance, it does not allow to go
beyond the tree level approximation, by taking into account in a consistent
way the effects of the higher order quantum corrections, which would provide
a better comparison with the results obtained from lattice numerical
simulations as well as from the studies of the Schwinger-Dyson equations.
The resolution of this important issue is due to Zwanziger who, by a
recursive characterization of the lowest eigenvalue of the Faddeev-Popov
operator, $-\partial _{\mu }D_{\mu }^{AB}=-\partial _{\mu }\left( \partial
_{\mu }\delta ^{AB}+gf^{ACB}A_{\mu }^{C}\right) $, has obtained a closed
expression for the nonlocal term which implements the restriction to the
Gribov region in the Landau gauge \cite{Zwanziger:1989mf,Zwanziger:1992qr}.
This term, known as the horizon function, is given by
\begin{equation}
S_{ZW}=-g^{2}\gamma ^{4}\int d^{4}x\;f^{ABC}A_{\mu }^{B}\left[ \left(
\partial _{\mu }D_{\mu }\right) ^{-1}\right] ^{AD}f^{DEC}A_{\mu }^{E}\;.
\label{i3}
\end{equation}
Notice that Zwanziger's horizon term, eq.$\left( \ref{i3}\right)
$, reduces to the Gribov term $\left( \ref{i1}\right) $ in the
quadratic approximation, in which the Faddeev-Popov operator,
$-\partial _{\mu }D_{\mu }^{AB}$ , is replaced by the Laplacian,
\textit{i.e.} by $-\delta ^{AB}\partial ^{2}$. To some extent,
Zwanziger's solution amounts to perform a re-summation of all
higher order terms neglected in Gribov's quadratic approximation.
In spite of its apparent nonlocality, the horizon function,
eq.$\left( \ref{i3}\right) $, can be cast in local form through
the introduction of a suitable set of additional fields.
Remarkably, the resulting local action turns out to be
multiplicatively renormalizable to all orders \cite
{Zwanziger:1989mf,Zwanziger:1992qr,Maggiore:1993wq}. This property
has made possible to evaluate the higher orders quantum
corrections by taking into account the effects of the restriction
to the Gribov region. In \cite
{Dudal:2005na} one finds a study at one loop order of the gluon condensate $%
\left\langle A_{\mu }^{A}A_{\mu }^{A}\right\rangle $ in the presence of the
horizon term $\left( \ref{i3}\right) $. More recently, the evaluation of the
two loop quantum corrections to the gluon and ghost propagators and their
relationship with the infrared freezing of the running coupling constant has
been worked out in \cite{Gracey:2005cx,Gracey:2006dr}. We emphasize that the
restriction to the Gribov region, as implemented by the Zwanziger horizon
function, gives us a useful theoretical framework in order to achieve a
rather clear understanding of the origin of the infrared suppression of the
gluon propagator as well as of the infrared enhancement of the ghost
propagator observed in the Landau gauge in both lattice numerical
simulations \cite
{Leinweber:1998uu,Bonnet:2001uh,Langfeld:2001cz,Cucchieri:2003di,Bloch:2003sk,Furui:2003jr,Silva:2004bv,Cucchieri:2004mf,Bogolubsky:2005wf}
and Schwinger-Dyson approach \cite
{vonSmekal:1997iss,vonSmekal:1997is,Atkinson:1997tu,Alkofer:2000wg,Watson:2001yv}%
. \newline
\newline
Concerning now the maximal Abelian gauge, till now, the restriction to the
Gribov region has been implemented within Gribov's quadratic approximation,
as reported in \cite{Capri:2005tj} in the case of $SU(2)$. For the analogue
of the nonlocal term $\left( \ref{i1}\right) $ we have obtained
\begin{equation}
S_{G}^{MAG}=-2\gamma ^{4}g^{2}\int d^{4}\!x\,A_{\mu }\frac{1}{\partial ^{2}}%
A_{\mu }\;,  \label{i4}
\end{equation}
where $A_{\mu }$ is the diagonal component of the gauge field, defined
according to the decomposition

\begin{equation}
\mathcal{A}_{\mu }=A_{\mu }^{a}T^{a}+A_{\mu }T^{3}\;,  \label{i5}
\end{equation}
where $T^{a}$, $a=1,2$, denote the off-diagonal generators of $SU(2)$, while
$T^{3}$ stands for the diagonal generator,
\begin{eqnarray}
\left[ T^{a},T^{b}\right] ~ &=&i~\varepsilon ^{ab}T^{3},  \nonumber \\
\left[ T^{3},T^{a}\right] ~ &=&i~\varepsilon ^{ab}T^{b},  \label{i6}
\end{eqnarray}
where
\begin{eqnarray}
\varepsilon ^{ab} &=&\varepsilon ^{ab3}\;,  \nonumber \\
\varepsilon ^{ab}\varepsilon ^{cd} &=&\delta ^{ac}\delta ^{bd}-\delta
^{ad}\delta ^{bc}\;.  \label{i7}
\end{eqnarray}
As in the case of the Landau gauge, the Gribov parameter $\gamma $ is not
free, being determined \cite{Capri:2005tj} by a gap equation\footnote{%
In the case of the maximal Abelian gauge, and for the gauge group $SU(2)$,
the gap equation defining the Gribov parameter $\gamma$ reads \cite
{Capri:2005tj}
\[
1=\frac{3}{4}g^{2}\int \frac{d^{4}k}{\left( 2\pi \right) ^{4}}\frac{1}{%
k^{4}+\gamma ^{4}}\;.
\]
} similar to eq.$\left( \ref{i2}\right) $. The addition of the Gribov term $%
\left( \ref{i4}\right) $ to the Yang-Mills action deeply modifies the
infrared behavior of the gluon and ghost propagators in the maximal Abelian
gauge. Let us mention here that the results which we have obtained \cite
{Capri:2005tj} for the diagonal and off-diagonal components of the gluon
propagator are in qualitative agreement with those reported in \cite
{Bornyakov:2003ee}, see next section for a brief review. Nevertheless, the
nonlocal term $\left( \ref{i4}\right) $ suffers from the same limitations of
the corresponding term, eq.$\left( \ref{i1}\right) $, of the Landau gauge,
requiring thus the construction of the analogue of Zwanziger's horizon term,
eq.$\left( \ref{i3}\right) $. This is the task of the present work. \newline
\newline
Before starting the description of the technical aspects, it might
be worth to give a short account of how the horizon term of the
maximal Abelian gauge has been identified. Having constructed this
term in the Gribov approximation, eq.$\left( \ref{i4}\right) $,
our strategy has been that of looking for a possible extension of
it, which enjoys the properties of \textit{localizability} and
\textit{renormalizability}. In other words, one
looks first at a possible nonlocal term which reduces to eq.$\left( \ref{i4}%
\right) $ in the quadratic approximation. Further, one requires that such
nonlocal term can be cast in local form by means of the introduction of a
suitable additional set of fields. Finally, the resulting local theory has
to be multiplicatively renormalizable. We underline that the requirements of
\textit{localizability} and \textit{renormalizability} are unavoidable in
order to have at our disposal a consistent computational framework. These
are strong requirements, which yield a very powerful criterion in order to
deal with nonlocal terms\footnote{%
See for instance the recent results \cite{Capri:2005dy,Capri:2006ne} on the
construction of a renormalizable nonabelian massive model based on the use
of the gauge invariant nonlocal operator $Tr \int d^4x F_{\mu\nu} \frac{1}{%
D^2}F_{\mu\nu}$.}. In the present case, a unique solution for the horizon
term in the maximal Abelian gauge, which reduces to expression $\left( \ref
{i4}\right) $\ and which fulfills the requirements of localizability and
renormalizability, has in fact emerged from our analysis. It reads
\begin{equation}
S_{\mathrm{Horizon}}=\gamma ^{4}g^{2}\int d^{4}\!x\,\varepsilon ^{ab}A_{\mu
}\left( \mathcal{M}^{-1}\right) ^{ac}\varepsilon ^{cb}A_{\mu }\ ,  \label{i8}
\end{equation}
where $\mathcal{M}^{ab}$ stands for the off-diagonal Faddeev-Popov operator
of the maximal Abelian gauge
\begin{equation}
\mathcal{M}^{ab}=-D_{\mu }^{ac}D_{\mu }^{cb}-g^{2}\varepsilon
^{ac}\varepsilon ^{bd}A_{\mu }^{c}A_{\mu }^{d}\;,  \label{19}
\end{equation}
with $D_{\mu }^{ab}$ being the covariant derivative with respect to the
diagonal component $A_{\mu }$,
\begin{equation}
D_{\mu }^{ab}=\delta ^{ab}\partial _{\mu }-g\varepsilon ^{ab}A_{\mu }\;.
\label{i10}
\end{equation}
Notice that the horizon term of eq.$\left( \ref{i8}\right) $ reduces
precisely to expression $\left( \ref{i4}\right) $ in the quadratic
approximation, amounting to replace the operator $\mathcal{M}^{ab}$ by the
Laplacian, \textit{i.e.} by $-\delta ^{ab}\partial ^{2}$. No other solution
for the horizon term in the maximal Abelian gauge, fulfilling the
requirements of localizability and renormalizability, has been found.
Although several candidates which reduce to expression $\left( \ref{i4}%
\right) $ can be easily written down, they are ruled out by the requirement
of localizability and renormalizabilty. Examples of such terms can be
obtained from $\left( \ref{i4}\right) $ by including more space-time
derivatives in the numerator, while considering higher powers of the
Faddeev-Popov operator $\mathcal{M}^{ab}$ in the denominator\footnote{%
An example of a term of this kind is given by the expression $\int
d^{4}\!x\,\varepsilon ^{ab}A_{\mu }\frac{\partial ^{2}}{\mathcal{M}^{ad}%
\mathcal{M}^{dc}}\varepsilon ^{cb}A_{\mu }$.}. However, the localization
procedure of such terms would require the introduction of dimensionless
auxiliary fields, a feature which jeopardizes the renormalizability of the
resulting action\footnote{%
Even if our present analysis has relied on a direct inspection of
possible candidates, it is worth mentioning that the requirements
of localizability and renormalizability might be linked in a
natural way to the deformation technique introduced in
\cite{Barnich:1993vg}, which could provide a more systematic way
to look at the construction of the horizon term for a generic
gauge, once the corresponding horizon function in the Gribov
approximation has been established. Essentially, this would amount
to start from the localized form of the horizon term in the Gribov
approximation, and search for the most general local deformation
which preserves all symmetries of the starting model as well as
its field content. By means of the use of the master equation and
of the antifield formalism, all possible deformations can be
identified in terms of cohomolgy classes of the nilpotent
functional operator associated to the master equation. The
construction of the extension of Gribov's horizon term through the
deformation approach is under investigation.}. Expression $\left(
\ref{i8}\right) $ can be seen thus as the minimal extension of the
nonlocal action $\left( \ref{i4}\right) $. It shares
great similarity with Zwanziger's horizon term in the Landau gauge, eq.$%
\left( \ref{i3}\right) $, as expressed by the presence of the
inverse of the Faddeev-Popov operator, $(\mathcal{M}^{-1})^{ab}$,
which looks rather natural. We remind in fact that the diagonal
ghost sector turns out to be unaffected by the presence of the
Gribov copies, see Appendix B of \cite{Capri:2005tj}. The proof of
the localizability and of the multiplicative renormalizability of
expression $\left( \ref{i8}\right) $ constitutes our main result.
\newline The present work is organized as follows. In Sect.2 we
give a brief account of the construction of the maximal Abelian
gauge fixing condition and of the results obtained in
\cite{Capri:2005tj} on the infrared behavior of the gluon and
ghost propagators. In Sect.3 we implement the localization
procedure for the nonlocal horizon term $\left( \ref{i8}\right) $,
presenting the $BRST\;$invariance as well as the rather rich set
of additional global symmetries of the resulting local action.
Sect.4 is devoted to a detailed proof of the multiplicative
renormalizability of the theory. The proof extends to all orders
by means of the use of the algebraic renormalization
\cite{Piguet:1995er}. An interesting feature of the theory is
represented by the existence of a residual local $U(1)$ Ward
identity, typical of the maximal Abelian
gauge. As a consequence, the nonrenormalization theorem, $Z_{g}Z_{A}^{1/2}=1$%
, relating the renormalization factors of the gauge coupling
constant $Z_{g}$ and of the diagonal gluon field $Z_{A}$, still
holds in the presence of the Gribov horizon. In Sect.5 we
introduce a generalized dimension two gluon operator and we prove
its multiplicative renormalizability in the presence of the
horizon term $\left( \ref{i8}\right) $. Furthermore, its anomalous
dimension is not an independent parameter of the theory, being
obtained from the renormalization factors of the gauge coupling
constant $g$ and of the diagonal anti-ghost field. Sect.6 contains
a few concluding remarks on further developments and on potential
applications for lattice numerical simulations.

\section{The gauge fixing condition for the maximal Abelian gauge}

In this section we introduce the gauge fixing condition for the maximal
Abelian gauge, providing a short account of the main results on the infrared
behavior of the gluon and ghost propagators obtained in \cite{Capri:2005tj}.
\newline
\newline
Similarly to the decomposition of the gauge field $\mathcal{A}_{\mu }$ given
in expression $\left( \ref{i5}\right) $, for the field strength one has
\begin{equation}
\mathcal{F}_{\mu \nu }=F_{\mu \nu }^{a}T^{a}+F_{\mu \nu }T^{3}\;,  \label{fs}
\end{equation}
with the off-diagonal and diagonal parts given by
\begin{eqnarray}
F_{\mu \nu }^{a} &=&D_{\mu }^{ab}A_{\nu }^{b}-D_{\nu }^{ab}A_{\mu }^{b}\;,
\label{fsc} \\
F_{\mu \nu } &=&\partial _{\mu }A_{\nu }-\partial _{\nu }A_{\mu
}+g\varepsilon ^{ab}A_{\mu }^{a}A_{\nu }^{b}\;,  \nonumber
\end{eqnarray}
where the covariant derivative $D_{\mu }^{ab}$ has been defined in eq.$%
\left( \ref{i10}\right) $. For the Yang-Mills action in Euclidean space one
obtains
\begin{equation}
S_{\mathrm{YM}}=\frac{1}{4}\int d^{4}x\,\left( F_{\mu \nu }^{a}F_{\mu \nu
}^{a}+F_{\mu \nu }F_{\mu \nu }\right) \;.  \label{ym}
\end{equation}
As it is easily checked, the classical action (\ref{ym}) is left invariant
by the gauge transformations
\begin{eqnarray}
\delta A_{\mu }^{a} &=&-D_{\mu }^{ab}{\omega }^{b}-g\varepsilon ^{ab}A_{\mu
}^{b}\omega \;,  \nonumber \\
\delta A_{\mu } &=&-\partial _{\mu }{\omega }-g\varepsilon ^{ab}A_{\mu
}^{a}\omega ^{b}\;.  \label{gauge}
\end{eqnarray}
The maximal Abelian gauge is obtained by demanding that the off-diagonal
components $A_{\mu }^{a}$ of the gauge field obey the nonlinear condition
\begin{equation}
D_{\mu }^{ab}A_{\mu }^{b}=0\;,  \label{offgauge}
\end{equation}
which follows by requiring that the auxiliary functional
\begin{equation}
\mathcal{R}[A]=\int {d^{4}x}A_{\mu }^{a}A_{\mu }^{a}\;,  \label{fmag}
\end{equation}
is stationary with respect to the gauge transformations (\ref{gauge}).
Moreover, as it is apparent from the presence of the covariant derivative $%
D_{\mu }^{ab}$, equation (\ref{offgauge}) allows for a residual local $U(1)$
invariance corresponding to the diagonal subgroup of $SU(2)$. This
additional invariance has to be fixed by means of a suitable gauge condition
on the diagonal component $A_{\mu }$, which will be chosen to be of the
Landau type, also adopted in lattice simulations, namely
\begin{equation}
\partial _{\mu }A_{\mu }=0\;.  \label{dgauge}
\end{equation}
The Faddeev-Popov operator, $\mathcal{M}^{ab}$, corresponding to the gauge
condition (\ref{offgauge}) is easily derived by taking the second variation
of the auxiliary functional $\mathcal{R}[A]$, being given by
\begin{equation}
\mathcal{M}^{ab}=-D_{\mu }^{ac}D_{\mu }^{cb}-g^{2}\varepsilon
^{ac}\varepsilon ^{bd}A_{\mu }^{c}A_{\mu }^{d}\;.  \label{offop}
\end{equation}
It enjoys the property of being Hermitian and, as pointed out in \cite
{Bruckmann:2000xd}, is the difference of two positive semidefinite operators
given, respectively, by $-D_{\mu }^{ac}D_{\mu }^{cb}$ and $g^{2}\varepsilon
^{ac}\varepsilon ^{bd}A_{\mu }^{c}A_{\mu }^{d}$. \newline
\newline
As discussed in \cite{Capri:2005tj}, in order to deal with the existence of
Gribov copies which affect the gauge condition (\ref{offgauge}), one
proceeds by restricting the domain of integration in the Feynman path
integral to the so called Gribov region $\mathcal{C}_{0}$, defined as the
set of fields fulfilling the gauge conditions (\ref{offgauge}), (\ref{dgauge}%
) and for which the Faddeev-Popov operator $\mathcal{M}^{ab}$ is positive
definite, namely
\begin{equation}
\mathcal{C}_{0}=\left\{ A_{\mu },\;A_{\mu }^{a},\;\partial _{\mu }A_{\mu
}=0,\;D_{\mu }^{ab}A_{\mu }^{b}=0,\;\mathcal{M}^{ab}=-D_{\mu }^{ac}D_{\mu
}^{cb}-g^{2}\varepsilon ^{ac}\varepsilon ^{bd}A_{\mu }^{c}A_{\mu
}^{d}>0\right\} \;.  \label{gr}
\end{equation}
The boundary, $\partial \mathcal{C}_{0}$, of the region $\mathcal{C}_{0}$,
where the first vanishing eigenvalue of $\mathcal{M}^{ab}$ appears, is
called the first Gribov horizon. The restriction of the domain of
integration to this region is supported by the possibility of generalizing
to the maximal Abelian gauge \cite{Capri:2005tj} Gribov's original result
\cite{Gribov:1977wm} stating that for any field located near a horizon there
is a gauge copy, close to the same horizon, located on the other side of the
horizon. \newline
\newline
The restriction to the region $\mathcal{C}_{0}$ can be implemented through
the no-pole condition on the off-diagonal two point ghost function. To the
first order, this condition amounts to the introduction of the nonlocal term
given in expression $\left( \ref{i4}\right) $. As a consequence, the tree
level gluon and ghost propagators get deeply modified in the infrared. More
precisely, both off-diagonal and diagonal transverse components of the gluon
propagator turn out to be suppressed in the infrared \cite{Capri:2005tj}.
The diagonal component of the gluon propagator is found to display the
characteristic Gribov type behavior, \textit{i.e.}
\begin{equation}
\left\langle A_{\mu }(k)A_{\nu }(-k)\right\rangle =\frac{k^{2}}{k^{4}+\gamma
^{4}}\left( \delta _{\mu \nu }-\frac{k_{\mu }k_{\nu }}{k^{2}}\right) \;.
\label{gl}
\end{equation}
The off-diagonal propagator turns out to be of the Yukawa type, being given
by
\begin{eqnarray}
\left\langle A_{\mu }^{a}(k)A_{\nu }^{b}(-k)\right\rangle &=&\delta ^{ab}%
\frac{1}{k^{2}+m^{2}}\left( \delta _{\mu \nu }-\frac{k_{\mu }k_{\nu }}{k^{2}}%
\right) \;,\;\;\;\;\;\;\;\;\;  \label{offg} \\
a,b &=&1,2  \nonumber
\end{eqnarray}
where $m$ denotes the off-diagonal dynamical mass originating from
the dimension two gluon condensate $\left\langle A_{\mu
}^{a}A_{\mu }^{a}\right\rangle $ \cite{Dudal:2004rx,Capri:2005tj}.
As already remarked, the behavior of the transverse diagonal and
off-diagonal gluon propagators, eqs.$\left( \ref{gl}\right)
,\left( \ref {offg}\right) $, is in qualitative agreement with the
lattice results \cite {Bornyakov:2003ee}. In the case of the ghost
propagator, it turns out that the off-diagonal component exhibits
infrared enhancement, according to
\begin{eqnarray}
\left. \mathcal{G}\left( k\right) \right| _{k=0} &\approx &\frac{\gamma ^{2}%
}{k^{4}}\;,  \label{offghg} \\
\mathcal{G}\left( k\right) &=&\frac{1}{2}\sum_{a}\left\langle \bar{c}%
^{a}(k)c^{a}(-k)\right\rangle \;,  \nonumber
\end{eqnarray}
where $(\bar{c}^{a},c^{a})\;$stand for the off-diagonal Faddeev-Popov
ghosts. Also, the diagonal component of the ghost propagator turns out to be
not affected by the restriction to the first horizon.

\section{Localization of the horizon function and symmetry content}

\subsection{Local action from the horizon}

In this section we describe the localization procedure for the
horizon term. Let us start by considering the partition function
of Yang-Mills theory quantized in the maximal Abelian gauge, in
the presence of the horizon term of eq.$\left( \ref{i8}\right) $,
namely
\begin{equation}
\mathcal{Z}=\int DA^{a}DADb^{a}DbD\bar{c}^{a}Dc^{a}D\bar{c}Dc\,\exp \left(
-\left( S_{\mathrm{YM}}+S_{\mathrm{MAG}}+S_{\mathrm{Horizon}}\right) \right)
\;,  \label{s31}
\end{equation}
where $S_{\mathrm{YM}}$ stands for the Yang-Mills action,
eq.(\ref{ym}), and $S_{\mathrm{MAG}}$ denotes the gauge fixing
term corresponding to the gauge conditions of
eqs.(\ref{offgauge}),(\ref{dgauge}). The fields $\left(
c^{a},\;\bar{c}^{a}\right) $ are the off-diagonal ghosts and
antighosts, while $\left( c,\;\bar{c}\right) $ denote the diagonal
ghost and antighost. Also, $\left( b^{a},\;b\right) $ are the
off-diagonal and diagonal Lagrange
multipliers enforcing conditions (\ref{offgauge}) and (\ref{dgauge}). $S_{%
\mathrm{MAG}}$ is given by the following expression \cite
{Fazio:2001rm,Dudal:2004rx}
\begin{equation}
S_{\mathrm{MAG}}=\int d^{4}\!x\,\left( b^{a}D_{\mu }^{ab}A_{\mu }^{b}-\bar{c}%
^{a}\mathcal{M}^{ab}c^{b}+g\varepsilon ^{ab}\bar{c}^{a}cD_{\mu }^{bc}A_{\mu
}^{c}+b\partial _{\mu }A_{\mu }+\bar{c}\partial _{\mu }(\partial _{\mu
}c+g\varepsilon ^{ab}A_{\mu }^{a}c^{b})\right) \;,  \label{s32}
\end{equation}
and $S_{\mathrm{Horizon}}$ is the horizon term $\left( \ref{i8}\right) $
\begin{equation}
S_{\mathrm{Horizon}}=\gamma ^{4}g^{2}\int d^{4}\!x\,\varepsilon ^{ab}A_{\mu
}\left( \mathcal{M}^{-1}\right) ^{ac}\varepsilon ^{cb}A_{\mu }\;.
\label{s33}
\end{equation}
\newline
As in the case of the Landau gauge \cite{Zwanziger:1989mf,Zwanziger:1992qr},
the horizon function (\ref{s33}) can be localized by means of a pair of
complex vector bosonic fields, $(\phi _{\mu }^{ab},\bar{\phi}_{\mu }^{ab}),$
$a,b=1,2$, according to
\begin{equation}
e^{-S_{\mathrm{Horizon}}}=\int D\bar{\phi}D\phi \,(\det \mathcal{M}%
)^{8}\,\exp \left( -\int d^{4}\!x\,\left( \bar{\phi}_{\mu }^{ab}\mathcal{M}%
^{ac}\phi _{\mu }^{cb}+\gamma ^{2}g\varepsilon ^{ab}(\phi _{\mu }^{ab}-\bar{%
\phi}_{\mu }^{ab})A_{\mu }\right) \right) \;,  \label{s34}
\end{equation}
\noindent where the determinant, $(\det \mathcal{M})^{8}$, takes into
account the Jacobian arising from the integration over the fields $(\phi
_{\mu }^{ab},\bar{\phi}_{\mu }^{ab})$. This term can also be localized by
means of a pair of vector anticommuting fields $(\omega _{\mu }^{ab},\bar{%
\omega}_{\mu }^{ab})$, namely
\begin{equation}
(\det \mathcal{M})^{8}=\int D\bar{\omega}D\omega \,\exp \left( \int
d^{4}\!x\,\bar{\omega}_{\mu }^{ab}\mathcal{M}^{ac}\omega _{\mu }^{cb}\right)
\;.  \label{s35}
\end{equation}
Therefore, we obtain a local action which reads
\begin{eqnarray}
\mathcal{Z} &=&\int D\Psi \,e^{-S_{\mathrm{Local}}}\;,  \nonumber \\
D\Psi &\equiv &DA^{a}DADb^{a}DbD\bar{c}^{a}Dc^{a}D\bar{c}DcD\bar{\phi}D\phi D%
\bar{\omega}D\omega \;,  \label{s36}
\end{eqnarray}
where
\begin{equation}
S_{\mathrm{Local}}=S_{\mathrm{YM}}+S_{\mathrm{MAG}}+S_{\phi \omega
}+S_{\gamma }\;,  \label{s37}
\end{equation}
with $S_{\phi \omega }$, $S_{\gamma }$ given by
\begin{eqnarray}
S_{\phi \omega } &=&\int d^{4}\!x\,\left( \bar{\phi}_{\mu }^{ab}\mathcal{M}%
^{ac}\phi _{\mu }^{cb}-\bar{\omega}_{\mu }^{ab}\mathcal{M}^{ac}\omega _{\mu
}^{cb}\right) \;,  \nonumber \\
&&  \nonumber \\
S_{\gamma } &=&\gamma ^{2}g\int d^{4}\!x\,\varepsilon ^{ab}\left( \phi _{\mu
}^{ab}-\bar{\phi}_{\mu }^{ab}\right) A_{\mu }\;.  \label{s38}
\end{eqnarray}

\subsection{BRST invariance}

In order to establish the $BRST\;$invariance of the resulting local theory,
we proceed as in \cite{Zwanziger:1989mf,Zwanziger:1992qr} and consider first
the particular case when $\gamma =0$, \textit{i.e.}
\begin{equation}
S_{\mathrm{Local}}|_{\gamma =0}=S_{\mathrm{YM}}+S_{\mathrm{MAG}}+S_{\phi
\omega }\;.  \label{gammazero}
\end{equation}
In this case we have in fact introduced nothing more than a unity written as
\begin{equation}
\int D\phi D\bar{\phi}D\omega D\bar{\omega}\,\exp \left( -\int
d^{4}\!x\,\left( \bar{\phi}_{\mu }^{ab}\mathcal{M}^{ac}\phi _{\mu }^{cb}-%
\bar{\omega}_{\mu }^{ab}\mathcal{M}^{ac}\omega _{\mu }^{cb}\right) \right)
=1\;.  \label{un}
\end{equation}
Furthermore, the action (\ref{gammazero}) may be written in a $BRST$
invariant fashion. To see this, let us introduce the following nilpotent $%
BRST $ transformations
\begin{eqnarray}
sA_{\mu }^{a} &=&-(D_{\mu }^{ab}c^{b}+g\varepsilon ^{ab}A_{\mu }^{b}c)\;,
\nonumber \\
sA_{\mu } &=&-(\partial _{\mu }c+g\varepsilon ^{ab}A_{\mu }^{a}c^{b})\;,
\nonumber \\
sc^{a} &=&g\varepsilon ^{ab}c^{b}c\;,  \nonumber \\
sc &=&\frac{g}{2}\varepsilon ^{ab}c^{a}c^{b}\;,  \nonumber \\
s\bar{c}^{a} &=&b^{a},\hspace{36.3pt}sb^{a}=0\;,  \nonumber \\
s\bar{c} &=&b,\hspace{46.5pt}sb=0\;,  \nonumber \\
s\phi _{\mu }^{ab} &=&\omega _{\mu }^{ab},\qquad s\omega _{\mu }^{ab}=0\;,
\nonumber \\
s\bar{\omega}_{\mu }^{ab} &=&\bar{\phi}_{\mu }^{ab},\hspace{25.1pt}s\bar{\phi%
}_{\mu }^{ab}=0\;,  \label{brst}
\end{eqnarray}
\begin{equation}
s^{2}=0\;.  \label{nilp}
\end{equation}
Now, let $S_{0}$ be the action defined by
\begin{equation}
S_{0}=S_{\mathrm{YM}}+s\int d^{4}\!x\,\left( \bar{c}^{a}D_{\mu }^{ab}A_{\mu
}^{b}+\bar{c}\partial _{\mu }A_{\mu }+\bar{\omega}_{\mu }^{ab}\mathcal{M}%
^{ac}\phi _{\mu }^{cb}\right) \;,  \label{so}
\end{equation}
which satisfies
\begin{equation}
sS_{0}=0\;.  \label{inso}
\end{equation}
Acting with the $BRST$ operator $s$, and recalling the expression of the
Faddeev-Popov operator $\mathcal{M}^{ab}$, eq.(\ref{offop}), we obtain
\begin{equation}
S_{0}=S_{\mathrm{YM}}+S_{\mathrm{MAG}}+S_{\phi \omega }+\int d^{4}\!x\,\bar{%
\omega}_{\mu }^{ab}\mathcal{F}^{ac}\phi _{\mu }^{cb}\;,  \label{actionzero}
\end{equation}
with
\begin{eqnarray}
\mathcal{F}^{ab} &=&2g\varepsilon ^{ac}(\partial _{\mu }c+g\varepsilon
^{de}A_{\mu }^{d}c^{e})D_{\mu }^{cb}+g\varepsilon ^{ab}\partial _{\mu
}(\partial _{\mu }c+g\varepsilon ^{cd}A_{\mu }^{c}c^{d})  \nonumber \\
&&-g^{2}(\varepsilon ^{ac}\varepsilon ^{bd}+\varepsilon ^{ad}\varepsilon
^{bc})A_{\mu }^{d}(D_{\mu }^{ce\!}c^{e}+g\varepsilon ^{ce}A_{\mu }^{e}c)\;.
\label{ftrem}
\end{eqnarray}
Expression (\ref{actionzero}) differs from the action (\ref{gammazero}) by
the presence of the last term. However, following \cite
{Zwanziger:1989mf,Zwanziger:1992qr}, we may transform $S_{\mathrm{Local}%
}|_{\gamma =0}$ into $S_{0}$ by performing the following shift in the
variable $\omega _{\mu }^{ab}$%
\begin{equation}
\omega _{\mu }^{ab}\to \omega _{\mu }^{ab}-\left( \mathcal{M}^{-1}\right)
^{ac}\mathcal{F}^{cd}\phi _{\mu }^{db}\;,  \label{shift}
\end{equation}
whose corresponding Jacobian turns out to be field independent.
Thus, the following equivalence holds, namely
\begin{equation}
\int D\Psi \,e^{-S_{0}} = \int D\Psi
\,e^{-S_{\mathrm{Local}}|_{\gamma =0}} \;. \label{npf}
\end{equation}
Further, let us consider the term $S_{\gamma }$ given by
(\ref{s38}). One can easily check that the term $\varepsilon
^{ab}A_{\mu }\bar{\phi}_{\mu }^{ab}$ which appears in $S_{\gamma
}$ can be written as
\begin{equation}
\varepsilon ^{ab}A_{\mu }\bar{\phi}_{\mu }^{ab}=\varepsilon ^{ab}s(A_{\mu }%
\bar{\omega}_{\mu }^{ab})-\varepsilon ^{ab}(\partial _{\mu }c+g\varepsilon
^{cd}A_{\mu }^{c}c^{d})\bar{\omega}_{\mu }^{ab}\;.  \label{strs}
\end{equation}
Once again, according to \cite{Zwanziger:1989mf,Zwanziger:1992qr}, we can
eliminate the last term of eq.(\ref{strs}) by means of the change of
variables
\begin{equation}
\omega _{\mu }^{ab}\to \omega _{\mu }^{ab}-\left(
\mathcal{M}^{-1}\right) ^{ac}\gamma ^{2}g\varepsilon
^{cb}(\partial _{\mu }c+g\varepsilon ^{de}A^{d}_{\mu}c^{e})\;.
\label{omsh}
\end{equation}
Therefore, for the partition function we obtain the final expression
\begin{equation}
\mathcal{Z}=\int D\Psi \,e^{-\left[ S_{0}+\gamma ^{2}g\int
d^{4}\!x\,\varepsilon ^{ab}\left(A_{\mu }\phi _{\mu }^{ab}-s\left(
A_{\mu }\bar{\omega}_{\mu }^{ab}\right) \right) \right] }\;.
\label{Phys}
\end{equation}
Notice also that, due to the identity
\begin{equation}
\gamma ^{2}g\int d^{4}\!x\,\varepsilon^{ab}\left( A_{\mu }\phi
_{\mu }^{ab}-s\left( A_{\mu }\bar{\omega}_{\mu }^{ab}\right)
\right) =\gamma ^{2}\int d^{4}\!x\,\left( D_{\mu }^{ab}\phi _{\mu
}^{ba}-s\left( D_{\mu }^{ab}\omega _{\mu }^{ba}\right) \right) \;,
\label{id}
\end{equation}
expression (\ref{Phys}) becomes
\begin{equation}
\mathcal{Z}=\int D\Psi \,e^{-\left[ S_{0}+\gamma ^{2}\int d^{4}\!x\,\left(
D_{\mu }^{ab}\phi _{\mu }^{ba}-s\left( D_{\mu }^{ab}\omega _{\mu
}^{ba}\right) \right) \right] }\;.  \label{pf}
\end{equation}
Nevertheless, due to the term $D_{\mu }^{ab}\phi _{\mu }^{ba}$, the action
\begin{equation}
S_{0}+\gamma ^{2}\int d^{4}\!x\,\left( D_{\mu }^{ab}\phi _{\mu
}^{ba}-s\left( D_{\mu }^{ab}\omega _{\mu }^{ba}\right) \right) \;,
\label{ninv}
\end{equation}
is not yet $BRST$ invariant, a point which can be dealt with by
means of the introduction of a pair of $BRST$ doublets of local
external sources \cite {Zwanziger:1989mf,Zwanziger:1992qr},
$\left( U_{\mu \nu }^{ab},M_{\mu \nu }^{ab}\right) $ and $\left(
V_{\mu \nu }^{ab},N_{\mu \nu }^{ab}\right) $, which transform as
\begin{eqnarray}
sU_{\mu \nu }^{ab} &=&-M_{\mu \nu }^{ab},\qquad sM_{\mu \nu }^{ab}=0\;,
\nonumber \\
sV_{\mu \nu }^{ab} &=&N_{\mu \nu }^{ab},\hspace{36.5pt}sN_{\mu \nu
}^{ab}=0\;.  \label{sou}
\end{eqnarray}
As pointed out in \cite{Zwanziger:1989mf,Zwanziger:1992qr}, the introduction
of these external sources allows us to promote expression (\ref{ninv}) to a $%
BRST\;$invariant action. In fact, let $S_{\mathrm{sources}}$ be the action
\begin{eqnarray}
S_{\mathrm{sources}} &=&s\int d^{4}\!x\,\left( -U_{\mu \nu
}^{ac}D_{\mu }^{ab}\phi _{\nu }^{bc}+V_{\mu \nu }^{ac}D_{\mu
}^{ab}\bar{\omega}_{\nu
}^{bc}\right)  \nonumber \\
&=&\int d^{4}\!x\,\left( M_{\mu \nu }^{ac}\,D_{\mu }^{ab}\phi _{\nu
}^{bc}+U_{\mu \nu }^{ac}\,s\left( D_{\mu }^{ab}\phi _{\nu }^{bc}\right)
+N_{\mu \nu }^{ac}\,D_{\mu }^{ab}\bar{\omega}_{\nu }^{bc}+V_{\mu \nu
}^{ac}\,s\left( D_{\mu }^{ab}\bar{\omega}_{\nu }^{bc}\right) \right) \;,
\label{acts}
\end{eqnarray}
which obviously satisfies
\begin{equation}
sS_{\mathrm{sources}}=0\;.  \label{finv}
\end{equation}
Moreover, when the sources $U$, $M$, $V$, $N$ attain their physical value
\cite{Zwanziger:1989mf,Zwanziger:1992qr}, defined by
\begin{eqnarray}
M_{\mu \nu }^{ab}|_{\mathrm{phys}} &=&-V_{\mu \nu }^{ab}|_{\mathrm{phys}%
}=\delta ^{ab}\delta _{\mu \nu }\gamma ^{2}\;,  \nonumber \\
U_{\mu \nu }^{ab}|_{\mathrm{phys}} &=&N_{\mu \nu }^{ab}|_{\mathrm{phys}}=0\;,
\label{physvalues}
\end{eqnarray}
it immediately follows that
\begin{equation}
S_{\mathrm{sources}}|_{\mathrm{phys}}=\int d^{4}\!x\,\left( \gamma
^{2}D_{\mu }^{ab}\phi _{\mu }^{ba}-\gamma ^{2}s\left( D_{\mu }^{ab}\bar{%
\omega}_{\mu }^{ba}\right) \right) \;,  \label{nice}
\end{equation}
which is nothing but expression (\ref{ninv}). One sees thus that the use of
the external sources $U$, $M$, $V$, $N$ enables us to introduce an extended
action $\Sigma _{0}$
\begin{equation}
\Sigma _{0}=S_{0}+S_{\mathrm{sources}}\;,  \label{extact}
\end{equation}
which enjoys the important property of being $BRST$ invariant,
\begin{equation}
s\Sigma _{0}=0\;,  \label{invb}
\end{equation}
while reducing to expression (\ref{ninv}) when the sources attain their
physical value, eq.(\ref{physvalues}). Explicitly, we have
\begin{eqnarray}
\Sigma _{0} &=&S_{YM}+S_{\mathrm{MAG}}+s\int d^{4}\!x\,\left( \bar{\omega}%
_{\mu }^{ab}\mathcal{M}^{ac}\phi _{\mu }^{cb}-U_{\mu \nu }^{ac}\,D_{\mu
}^{ab}\phi _{\nu }^{bc}+V_{\mu \nu }^{ac}\,D_{\mu }^{ab}\bar{\omega}_{\nu
}^{bc}\right)  \nonumber \\
&=&S_{\mathrm{YM}}+S_{\mathrm{MAG}}+\int d^{4}\!x\left( \,\bar{\phi}_{\mu
}^{ab}\mathcal{M}^{ac}\phi _{\mu }^{cb}-\bar{\omega}_{\mu }^{ab}\mathcal{M}%
^{ac}\omega _{\mu }^{cb}+\bar{\omega}_{\mu }^{ab}\mathcal{F}^{ac}\phi _{\mu
}^{cb}+M_{\mu \nu }^{ac}\,D_{\mu }^{ab}\phi _{\nu }^{bc}+N_{\mu \nu
}^{ac}\,D_{\mu }^{ab}\bar{\omega}_{\nu }^{bc}\right.  \nonumber \\
&&\left. +U_{\mu \nu }^{ac}[D_{\mu }^{ab}\omega _{\nu }^{bc}+g\varepsilon
^{ab}(\partial _{\mu }c+g\varepsilon ^{de}A_{\mu }^{d}c^{e})\phi _{\nu
}^{bc}]+V_{\mu \nu }^{ac}[D_{\mu }^{ab}\bar{\phi}_{\nu }^{bc}+g\varepsilon
^{ab}(\partial _{\mu }c+g\varepsilon ^{de}A_{\mu }^{d}c^{e})\bar{\omega}%
_{\nu }^{bc}]\right) \;.  \nonumber \\
&&  \label{Sigma0}
\end{eqnarray}

\subsection{Inclusion of the quartic ghost term}

Although being $BRST\;$invariant, the action $\Sigma _{0}$ is not
yet the most general classical action to start with. As discussed
in previous works \cite{Fazio:2001rm, Dudal:2004rx}, the
nonlinearity of the gauge condition (\ref{offgauge}) requires the
introduction of a quartic term in the Faddeev-Popov ghost fields
\begin{equation}
\frac{\alpha }{2}g^{2}\overline{c}^{a}c^{a}\overline{c}^{b}c^{b}\;,
\label{qgh}
\end{equation}
which is in fact needed for renormalizability purposes. The parameter $%
\alpha $ in expression (\ref{qgh}) is a gauge parameter. In our
case, due to the presence of the localizing fields $(\phi _{\mu
}^{ab},\bar{\phi}_{\mu }^{ab},\omega _{\mu
}^{ab},\bar{\omega}_{\mu }^{ab})$, the quartic ghost term
(\ref{qgh}) is introduced in a $BRST\;$invariant way through the
following
action $\Sigma _{\alpha }$%
\begin{eqnarray}
\Sigma _{\alpha } &=&s\int d^{4}x\frac{\alpha }{2}\left( \overline{c}%
^{a}b^{a}-g\varepsilon ^{ab}\overline{c}^{a}\overline{c}^{b}c+g^{2}\bar{%
\omega}_{\mu }^{ac}\phi _{\mu }^{ac}\left( \bar{\phi}_{\nu }^{bd}\phi _{\nu
}^{bd}-\bar{\omega}_{\nu }^{bd}\omega _{\nu }^{bd}\right) -2g^{2}\bar{\omega}%
_{\mu }^{ac}\phi _{\mu }^{ac}\overline{c}^{b}c^{b}\right)  \nonumber \\
&=&\frac{\alpha }{2}\int d^{4}x\left( b^{a}b^{a}-2g\varepsilon ^{ab}b^{a}%
\overline{c}^{b}c+g^{2}\overline{c}^{a}c^{a}\overline{c}^{b}c^{b}+g^{2}%
\left( \bar{\phi}_{\mu }^{ac}\phi _{\mu }^{ac}-\bar{\omega}_{\mu
}^{ac}\omega _{\mu }^{ac}\right) \left( \bar{\phi}_{\nu }^{bd}\phi _{\nu
}^{bd}-\bar{\omega}_{\nu }^{bd}\omega _{\nu }^{bd}\right) \right.  \nonumber
\\
&&\;\;\;\;\;\;\;\;\;\;\left. -2g^{2}\left( \bar{\phi}_{\mu }^{ac}\phi _{\mu
}^{ac}-\bar{\omega}_{\mu }^{ac}\omega _{\mu }^{ac}\right) \overline{c}%
^{b}c^{b}+2g^{2}\bar{\omega}_{\mu }^{ac}\phi _{\mu }^{ac}b^{b}c^{b}-2g^{3}%
\bar{\omega}_{\mu }^{ad}\phi _{\mu }^{ad}\varepsilon ^{bc}\overline{c}%
^{b}c^{c}c\right) \;.  \label{salph}
\end{eqnarray}
Adding the $BRST$ invariant term $\Sigma _{\alpha }$ to the action
$\Sigma _{0}$, modifies the equation of motion of the off-diagonal
Lagrange multiplier $b^{a}$, according to
\begin{equation}
\frac{\delta \left( \Sigma _{0}+\Sigma _{\alpha }\right) }{\delta b^{a}}%
=D_{\mu }^{ab}A_{\mu }^{b}+\alpha \left( b^{a}-g\varepsilon ^{ab}\overline{c}%
^{b}c+g^{2}\bar{\omega}_{\mu }^{bc}\phi _{\mu }^{bc}c^{a}\right) \;.
\label{bem}
\end{equation}
Therefore, we see that maximal Abelian gauge condition
(\ref{offgauge}), $D_{\mu }^{ab}A_{\mu }^{b}=0$, is attained in
the limit $\alpha \rightarrow 0 $, which has to be performed after
the removal of the ultraviolet divergences \cite{Fazio:2001rm,
Dudal:2004rx}. We also remark that expression (\ref{salph})
contains a unique free parameter, namely, the gauge parameter
$\alpha $. As a consequence, the whole term $\Sigma _{\alpha }$
vanishes in the limit $\alpha \rightarrow 0$, allowing us to
integrate out the auxiliary fields $(\phi _{\mu
}^{ab},\bar{\phi}_{\mu }^{ab},\omega _{\mu
}^{ab},\bar{\omega}_{\mu }^{ab})$, thus recovering the horizon
term $\left( \ref{i8}\right) $. As we shall see in the next
section, this important feature follows from the fact that the
action $\left( \Sigma _{0}+\Sigma _{\alpha }\right) $ fulfills
several Ward identities which, in particular, uniquely fix the
form of the term $\Sigma _{\alpha }$.

\subsection{The global $U(8)$ symmetry}

In addition to the $BRST$ invariance, and in complete analogy with
the case of
the Landau gauge \cite{Zwanziger:1989mf,Zwanziger:1992qr}, the action $%
\left( \Sigma _{0}+\Sigma _{\alpha }\right) $ displays a global symmetry $%
U(f)$, $f=8$, expressed by
\begin{equation}
\mathcal{Q}_{\mu \nu }^{ab}\left( \Sigma _{0}+\Sigma _{\alpha }\right) =0\;,
\label{uf}
\end{equation}
with
\begin{eqnarray}
\mathcal{Q}_{\mu \nu }^{ab} &=&\int d^{4}\!x\,\left( \phi _{\mu }^{ca}\frac{%
\delta }{\delta \phi _{\nu }^{cb}}-\bar{\phi}_{\nu }^{cb}\frac{\delta }{%
\delta \bar{\phi}_{\mu }^{ca}}+\omega _{\mu }^{ca}\frac{\delta }{\delta
\omega _{\nu }^{cb}}-\bar{\omega}_{\nu }^{cb}\frac{\delta }{\delta \bar{%
\omega}_{\mu }^{ca}}\right.  \nonumber \\
&&\left. +V_{\sigma \mu }^{ca}\frac{\delta }{\delta V_{\sigma \nu }^{cb}}%
-M_{\sigma \nu }^{cb}\frac{\delta }{\delta M_{\sigma \mu }^{ca}}+N_{\sigma
\mu }^{ca}\frac{\delta }{\delta N_{\sigma \nu }^{cb}}-U_{\sigma \nu }^{cb}%
\frac{\delta }{\delta U_{\sigma \mu }^{ca}}\right) \;.  \label{ufop}
\end{eqnarray}
The presence of the global invariance $U(8)$ means that one can
make use \cite{Zwanziger:1989mf,Zwanziger:1992qr} of the composite
index $i\equiv (a,\mu )$, $i=1,\dots ,8$. Therefore, setting
\begin{equation}
(\phi _{\mu }^{ab},\bar{\phi}_{\mu }^{ab},\omega _{\mu }^{ab},\bar{\omega}%
_{\mu }^{ab})=(\phi _{i}^{a},\bar{\phi}_{i}^{a},\omega _{i}^{a},\bar{\omega}%
_{i}^{a})\;,  \label{if}
\end{equation}
and
\begin{equation}
(U_{\mu \nu }^{ab},V_{\mu \nu }^{ab},M_{\mu \nu }^{ab},N_{\mu \nu
}^{ab})=(U_{\mu i}^{a},V_{\mu i}^{a},M_{\mu i}^{a},N_{\mu i}^{a})\;,
\label{is}
\end{equation}
we rewrite expression (\ref{extact}) as
\begin{eqnarray}
\Sigma _{0}+\Sigma _{\alpha } &=&S_{\mathrm{YM}}+S_{\mathrm{MAG}}+\int
d^{4}\!x\,\left( \bar{\phi}_{i}^{a}\mathcal{M}^{ab}\phi _{i}^{b}-\bar{\omega}%
_{i}^{a}\mathcal{M}^{ab}\omega _{i}^{b}+\bar{\omega}_{i}^{a}\mathcal{F}%
^{ab}\phi _{i}^{b}+M_{\mu i}^{a}\,D_{\mu }^{ab}\phi _{i}^{b}+N_{\mu
i}^{a}\,D_{\mu }^{ab}\bar{\omega}_{i}^{b}\right.  \nonumber \\
&&\left. +U_{\mu i}^{a}[D_{\mu }^{ab}\omega _{i}^{b}+g\varepsilon
^{ab}(\partial _{\mu }c+g\varepsilon ^{cd}A_{\mu }^{c}c^{d})\phi
_{i}^{b}]+V_{\mu i}^{a}[D_{\mu }^{ab}\bar{\phi}_{i}^{b}+g\varepsilon
^{ab}(\partial _{\mu }c+g\varepsilon ^{cd}A_{\mu }^{c}c^{d})\bar{\omega}%
_{i}^{b}]\;\right) \;  \nonumber \\
&&+\frac{\alpha }{2}\int d^{4}x\left( b^{a}b^{a}-2g\varepsilon ^{ab}b^{a}%
\overline{c}^{b}c+g^{2}\overline{c}^{a}c^{a}\overline{c}^{b}c^{b}+g^{2}%
\left( \bar{\phi}_{i}^{a}\phi _{i}^{a}-\bar{\omega}_{i}^{a}\omega
_{i}^{a}\right) \left( \bar{\phi}_{j}^{b}\phi _{j}^{b}-\bar{\omega}%
_{j}^{b}\omega _{j}^{b}\right) \right.  \nonumber \\
&&\;\left. -2g^{2}\left( \bar{\phi}_{i}^{a}\phi _{i}^{a}-\bar{\omega}%
_{i}^{a}\omega _{i}^{a}\right) \overline{c}^{b}c^{b}+2g^{2}\bar{\omega}%
_{i}^{a}\phi _{i}^{a}b^{b}c^{b}-2g^{3}\bar{\omega}_{i}^{a}\phi
_{i}^{a}\varepsilon ^{bc}\overline{c}^{b}c^{c}c\right)  \label{S0}
\end{eqnarray}
For the symmetry generator we have
\begin{eqnarray}
\mathcal{Q}_{ij} &=&\int d^{4}\!x\,\left( \phi _{i}^{a}\frac{\delta }{\delta
\phi _{j}^{a}}-\bar{\phi}_{j}^{a}\frac{\delta }{\delta \bar{\phi}_{i}^{a}}%
+\omega _{i}^{a}\frac{\delta }{\delta \omega _{j}^{a}}-\bar{\omega}_{j}^{a}%
\frac{\delta }{\delta \bar{\omega}_{i}^{a}}\right.  \nonumber \\
&&\left. +V_{\mu i}^{a}\frac{\delta }{\delta V_{\mu j}^{a}}-M_{\mu j}^{a}%
\frac{\delta }{\delta M_{\mu i}^{a}}+N_{\mu i}^{a}\frac{\delta }{\delta
N_{\mu j}^{a}}-U_{\mu j}^{a}\frac{\delta }{\delta U_{\mu i}^{a}}\right) \;.
\label{qif}
\end{eqnarray}
By means of the trace of the operator $\mathcal{Q}_{ij}$, \textit{i.e.}, $%
\mathcal{Q}_{ii}\equiv \mathcal{Q}_{f}$, the $i$-valued fields turn out to
possess an additional quantum number, displayed in Table $\left( \ref{table1}%
\right) $, together with the dimension and the ghost number.
\begin{table}[t]
\centering
\begin{tabular}{lcccccccccccc}
\hline
& $A$ & $b$ & $\bar c$ & $c$ & $\phi$ & $\bar\phi$ & $\omega$ & $\bar\omega$
& $U$ & $V$ & $M$ & $N\phantom{\Bigl|}$ \\ \hline\hline
dimension & $1$ & $2$ & $2$ & $0$ & $1$ & $1$ & $1$ & $1$ & $2$ & $2$ & $2$
& $2$ \\
gh number & $0$ & $0$ & $-1$ & $1$ & $0$ & $0$ & $1$ & $-1$ & $-1$ & $0$ & $%
0 $ & $1$ \\
$\mathcal{Q}_f$-charge & $0$ & $0$ & $0$ & $0$ & $1$ & $-1$ & $1$ & $-1$ & $%
-1$ & $1$ & $-1$ & $1$ \\ \hline
\end{tabular}
\caption{Quantum numbers of the fields and sources}
\label{table1}
\end{table}

\section{Identification of the final complete classical action $\Sigma $}

Let us proceed by establishing the rich set of Ward identities
which will enable us to analyse the renormalizability of the
theory to all orders. Let us first identify the final complete
action to start with. To this purpose, we have
to properly define the composite field operators entering the $BRST\;$%
transformations $\left( \ref{brst}\right) $,$\left(
\ref{sou}\right) $. We notice that the $BRST$ transformation of
the gauge field $A_{\mu }^{a}$ can be written as the sum of two
composite operators, \textit{i.e.}
\begin{equation}
sA_{\mu }^{a}=\mathcal{O}_{1}+\mathcal{O}_{2}  \label{2op}
\end{equation}
where
\begin{equation}
\mathcal{O}_{1}=-D_{\mu }^{ab}c^{b},\qquad \mathcal{O}_{2}=-g\varepsilon
^{ab}A_{\mu }^{b}c\;.  \label{2opd}
\end{equation}
Moreover, thanks to the nilpotency of the $BRST$ operator, $s^{2}=0$, it
follows that
\begin{equation}
s\mathcal{O}_{1}=-s\mathcal{O}_{2}\;.  \label{s2op}
\end{equation}
Therefore, the two composite operators,
$\mathcal{O}_{1},\mathcal{O}_{2}$, can be defined by means of the
introduction of the external sources $\left( \Omega _{\mu
}^{a},\tau _{\mu
}^{a},\xi _{\mu }^{a}\right) $%
\begin{equation}
\Sigma _{\mathrm{ext}}^{(1)}=\int d^{4}\!x\,\left( \Omega _{\mu }^{a}\left(
-D_{\mu }^{ab}c^{b}\right) +\tau _{\mu }^{a}\left( -g\varepsilon ^{ab}A_{\mu
}c\right) +\xi _{\mu }^{a}\,s\!\left( -g\varepsilon ^{ab}A_{\mu }c\right)
\right) \;.  \label{2opsou}
\end{equation}
To guarantee the $BRST$ invariance of $\Sigma
_{\mathrm{ext}}^{(1)}$, we require that
\begin{eqnarray}
s\xi _{\mu }^{a} &=&-(\Omega _{\mu }^{a}-\tau _{\mu }^{a})\;,  \nonumber \\
s\Omega _{\mu }^{a} &=&s\tau _{\mu }^{a}=0\;.  \label{s2s}
\end{eqnarray}
Moreover, it is easily checked that the new action
\begin{equation}
\Sigma _{0}+\Sigma _{\alpha }+\Sigma _{\mathrm{ext}}^{(1)}\;,  \label{S1}
\end{equation}
is left invariant by the following set of transformations:

\begin{itemize}
\item  {the $\delta _{i}$ symmetry}
\end{itemize}
\begin{eqnarray}
\delta _{i}\bar{c}^{a} &=&\phi _{i}^{a}\;,  \nonumber \\
\delta _{i}\bar{\phi}_{j}^{a} &=&\delta _{ij}c^{a}\;,  \nonumber \\
\delta _{i}b^{a} &=&g\varepsilon ^{ab}\phi _{i}^{b}c\;,  \nonumber \\
\delta _{i}\Omega _{\mu }^{a} &=&V_{\mu i}^{a}\;,  \label{di}
\end{eqnarray}
\begin{itemize}
\item  {the $\tilde{\delta}_{i}$ symmetry}
\end{itemize}
\begin{eqnarray}
\tilde{\delta}_{i}\bar{c}^{a} &=&\bar{\omega}_{i}^{a}\;,  \nonumber \\
\tilde{\delta}_{i}\omega _{j}^{a} &=&-\delta _{ij}c^{a}\;,  \nonumber \\
\tilde{\delta}_{i}b^{a} &=&g\varepsilon ^{ab}\bar{\omega}_{i}^{b}c\;,
\nonumber \\
\tilde{\delta}_{i}\Omega _{\mu }^{a} &=&-U_{\mu i}^{a}\;,  \label{dti}
\end{eqnarray}
with
\begin{equation}
\delta _{i}\left( \Sigma _{0}+\Sigma _{\alpha }+\Sigma _{\mathrm{ext}%
}^{(1)}\right) =\tilde{\delta}_{i}\left( \Sigma _{0}+\Sigma _{\alpha
}+\Sigma _{\mathrm{ext}}^{(1)}\right) =0\;.  \label{ddt}
\end{equation}
As transformations $\left( \text{\ref{di}}\right) ,$ $\left( \text{\ref{dti}}%
\right) $ contain composite field operators, {\it i.e.}
$g\varepsilon ^{ab}\phi _{i}^{b}c$ and $g\varepsilon
^{ab}\bar{\omega}_{i}^{b}c$, we define them by means of external
sources $\left( \eta _{i}^{a},\lambda _{i}^{a}\right) $ and
$\left( \vartheta _{i}^{a},\rho _{i}^{a}\right) $, giving rise to
two set of $BRST$ doublets
\begin{eqnarray}
s\eta _{i}^{a} &=&\lambda _{i}^{a},\qquad s\lambda _{i}^{a}=0\;,  \nonumber
\\
s\vartheta _{i}^{a} &=&\rho _{i}^{a},\qquad s\rho _{i}^{a}=0\;,
\label{2doub}
\end{eqnarray}
so that
\begin{eqnarray}
\Sigma _{\mathrm{ext}}^{(2)} &=&s\int d^{4}\!x\,g\varepsilon ^{ab}\left(
\eta _{i}^{a}\phi _{i}^{b}c+\vartheta _{i}^{a}\bar{\omega}_{i}^{b}c\right)
\nonumber \\
&=&\int d^{4}\!x\,\left( g\varepsilon ^{ab}\lambda _{i}^{a}\phi
_{i}^{b}c+\eta _{i}^{a}[g\varepsilon ^{ab}\omega _{i}^{b}c+\frac{g^{2}}{2}%
\varepsilon ^{ab}\varepsilon ^{cd}\phi _{i}^{b}c^{c}c^{d}]\right.  \nonumber
\\
&&\left. +g\varepsilon ^{ab}\rho _{i}^{a}\bar{\omega}_{i}^{b}c-\vartheta
_{i}^{a}[g\varepsilon ^{ab}\bar{\phi}_{i}^{b}c-\frac{g^{2}}{2}\varepsilon
^{ab}\varepsilon ^{cd}\bar{\omega}_{i}^{b}c^{c}c^{d}]\;\right) \;.
\label{ext2}
\end{eqnarray}
Further, introducing the $BRST\;$invariant sources $\left( \Omega _{\mu
},L^{a},L\right) $, coupled respectively to the nonlinear $BRST\;$%
transformations of the fields $A_{\mu }$, $c^{a}$, $c$, for the
final expression of the complete action $\Sigma$ we shall start
with, we have
\begin{equation}
\Sigma =\Sigma _{0}+\Sigma _{\alpha }+\Sigma _{\mathrm{ext}}^{(1)}+\Sigma _{%
\mathrm{ext}}^{(2)}+s\int d^{4}\!x\,\left( -\Omega _{\mu }A_{\mu
}+L^{a}c^{a}+Lc-\chi \,U_{\mu i}^{a}V_{\mu i}^{a}\right) \;,  \label{fact}
\end{equation}
namely
\begin{eqnarray}
\Sigma &=&S_{\mathrm{YM}}+S_{\mathrm{MAG}}+s\int d^{4}\!x\,\left( \bar{\omega%
}_{i}^{a}{\mathcal{M}}^{ab}\phi _{i}^{b}-U_{\mu i}^{a}\,D_{\mu }^{ab}\!\phi
_{i}^{b}+V_{\mu i}^{a}\,D_{\mu }^{ab}\!\bar{\omega}_{i}^{b}+g\varepsilon
^{ab}\eta _{i}^{a}\phi _{i}^{b}c+g\varepsilon ^{ab}\vartheta _{i}^{a}\bar{%
\omega}_{i}^{b}c\right.  \nonumber \\
\; &&\left. -\Omega _{\mu }^{a}A_{\mu }^{a}-g\varepsilon ^{ab}\xi _{\mu
}^{a}A_{\mu }^{b}c-\Omega _{\mu }A_{\mu }+L^{a}c^{a}+Lc-\chi \,U_{\mu
i}^{a}V_{\mu i}^{a}\right) \;  \nonumber \\
&&+s\int d^{4}x\frac{\alpha }{2}\left( \overline{c}^{a}b^{a}-g\varepsilon
^{ab}\overline{c}^{a}\overline{c}^{b}c+g^{2}\bar{\omega}_{i}^{a}\phi
_{i}^{a}\left( \bar{\phi}_{j}^{b}\phi _{j}^{b}-\bar{\omega}_{j}^{b}\omega
_{j}^{b}\right) -2g^{2}\bar{\omega}_{i}^{a}\phi _{i}^{a}\overline{c}%
^{b}c^{b}\right)  \nonumber \\
&=&S_{\mathrm{YM}}+\int d^{4}\!x\,\left( b^{a}\,D_{\mu }^{ab\!}A_{\mu }^{b}-%
\bar{c}^{a}\mathcal{M}^{ab}c^{b}+g\varepsilon ^{ab}\bar{c}^{a}c\,D_{\mu
}^{bc\!}A_{\mu }^{c}+b\,\partial _{\mu \!}A_{\mu }+\bar{c}\partial _{\mu
}(\partial _{\mu }c+g\varepsilon ^{ab}A_{\mu }^{a}c^{b})\right.  \nonumber \\
&&+\bar{\phi}_{i}^{a}\mathcal{M}^{ab}\phi _{i}^{b}-\bar{\omega}_{i}^{a}%
\mathcal{M}^{ab}\omega _{i}^{b}+\bar{\omega}_{i}^{a}\mathcal{F}^{ab}\phi
_{i}^{b}+M_{\mu i}^{a}\,D_{\mu }^{ab\!}\phi _{i}^{b}+U_{\mu i}^{a}[D_{\mu
}^{ab\!}\omega _{i}^{b}+g\varepsilon ^{ab}(\partial _{\mu }c+g\varepsilon
^{cd}A_{\mu }^{c}c^{d})\phi _{i}^{b}]  \nonumber \\
&&+N_{\mu i}^{a}\,D_{\mu }^{ab\!}\bar{\omega}_{i}^{b}+V_{\mu i}^{a}[D_{\mu
}^{ab\!}\bar{\phi}_{i}^{b}+g\varepsilon ^{ab}(\partial _{\mu }c+g\varepsilon
^{cd}A_{\mu }^{c}c^{d})\bar{\omega}_{i}^{b}]-\Omega _{\mu }^{a}\,D_{\mu
}^{ab\!}c^{b}-g\varepsilon ^{ab}\tau _{\mu }^{a}A_{\mu }^{b}c  \nonumber \\
&&+\xi _{\mu }^{a}[g\varepsilon ^{ab}(D_{\mu }^{bc\!}c^{c})c-\frac{g^{2}}{2}%
\varepsilon ^{ab}\varepsilon ^{cd}A_{\mu }^{b}c^{c}c^{d}]-\Omega _{\mu
}(\partial _{\mu }c+g\varepsilon ^{ab}A_{\mu }^{a}c^{b})+g\varepsilon
^{ab}L^{a}c^{b}c+\frac{g}{2}\varepsilon ^{ab}Lc^{a}c^{b}  \nonumber \\
&&+g\varepsilon ^{ab}\lambda _{i}^{a}\phi _{i}^{b}c+\eta
_{i}^{a}[g\varepsilon ^{ab}\omega _{i}^{b}c+\frac{g^{2}}{2}\varepsilon
^{ab}\varepsilon ^{cd}\phi _{i}^{b}c^{c}c^{d}]+g\varepsilon ^{ab}\rho
_{i}^{a}\bar{\omega}_{i}^{b}c-\vartheta _{i}^{a}[g\varepsilon ^{ab}\bar{\phi}%
_{i}^{b}c-\frac{g^{2}}{2}\varepsilon ^{ab}\varepsilon ^{cd}\bar{\omega}%
_{i}^{b}c^{c}c^{d}]  \nonumber \\
&&\left. {\ }{\ }{\ }{\ }{\ }{\ }{\ }{\ }{\ }{\ }+\chi (M_{\mu
i}^{a}V_{\mu i}^{a}+U_{\mu i}^{a}N_{\mu
i}^{a})\;\right) \;  \nonumber \\
&&+\frac{\alpha }{2}\int d^{4}x\left( b^{a}b^{a}-2g\varepsilon ^{ab}b^{a}%
\overline{c}^{b}c+g^{2}\overline{c}^{a}c^{a}\overline{c}^{b}c^{b}+g^{2}%
\left( \bar{\phi}_{i}^{a}\phi _{i}^{a}-\bar{\omega}_{i}^{a}\omega
_{i}^{a}\right) \left( \bar{\phi}_{j}^{b}\phi _{j}^{b}-\bar{\omega}%
_{j}^{b}\omega _{j}^{b}\right) \right.  \nonumber \\
&&\;\left. {\ }{\ }{\ }{\ }{\ }{\ }{\ }{\ }{\ }{\ } -2g^{2}\left( \bar{\phi}_{i}^{a}\phi _{i}^{a}-\bar{\omega}%
_{i}^{a}\omega _{i}^{a}\right) \overline{c}^{b}c^{b}+2g^{2}\bar{\omega}%
_{i}^{a}\phi _{i}^{a}b^{b}c^{b}-2g^{3}\bar{\omega}_{i}^{a}\phi
_{i}^{a}\varepsilon ^{bc}\overline{c}^{b}c^{c}c\right) \;.
\label{sigma}
\end{eqnarray}
One should notice that the term quadratic in the sources $\left(
M_{\mu i}^{a},V_{\mu i}^{a},U_{\mu i}^{a},N_{\mu i}^{a}\right) $,
{\it i.e.} $\chi (M_{\mu i}^{a}V_{\mu i}^{a}+U_{\mu i}^{a}N_{\mu
i}^{a})$, in eq.$\left( \text{\ref{sigma}}\right) $ is of
dimension four. As such, it is allowed by power counting and has
to be added for renormalizability purposes. The parameter $\chi $
stands for a free coefficient. \newline
\newline
Expression $\left( \text{\ref{sigma}}%
\right) $ represents our starting action.

\subsection{Ward identities}

It turns out that the classical action $\Sigma $, eq.$\left( \text{\ref{sigma}}%
\right) $, obeys the following set of Ward identities:
\begin{itemize}
\item  the Slavnov-Taylor identity
\begin{equation}
\mathcal{S}(\Sigma )=0\;,  \label{st}
\end{equation}
with
\end{itemize}
\begin{eqnarray}
\mathcal{S}(\Sigma ) &\equiv&\int d^{4}\!x\left( \,\left( \frac{\delta \Sigma }{%
\delta \Omega _{\mu }^{a}}+\frac{\delta \Sigma }{\delta \tau _{\mu }^{a}}%
\right) \frac{\delta \Sigma }{\delta A_{\mu }^{a}}+\frac{\delta \Sigma }{%
\delta \Omega _{\mu }}\frac{\delta \Sigma }{\delta A_{\mu }}+\frac{\delta
\Sigma }{\delta L^{a}}\frac{\delta \Sigma }{\delta c^{a}}+\frac{\delta
\Sigma }{\delta L}\frac{\delta \Sigma }{\delta c}+b^{a}\frac{\delta \Sigma }{%
\delta \bar{c}^{a}}+b\frac{\delta \Sigma }{\delta c}\right.  \nonumber \\
&&\left. +\omega _{i}^{a}\frac{\delta \Sigma }{\delta \phi _{i}^{a}}+\bar{%
\phi}_{i}^{a}\frac{\delta \Sigma }{\delta \bar{\omega}_{i}^{a}}+N_{\mu i}^{a}%
\frac{\delta \Sigma }{\delta V_{\mu i}^{a}}-M_{\mu i}^{a}\frac{\delta \Sigma
}{\delta U_{\mu i}^{a}}-(\Omega _{\mu }^{a}-\tau _{\mu }^{a})\frac{\delta
\Sigma }{\delta \xi _{\mu }^{a}}+\lambda _{i}^{a}\frac{\delta \Sigma }{%
\delta \eta _{i}^{a}}+\rho _{i}^{a}\frac{\delta \Sigma }{\delta \vartheta
_{i}^{a}}\right) \;,  \nonumber \\
&&  \label{ste}
\end{eqnarray}
\begin{itemize}
\item  the Ward identities corresponding to the $\delta ${$_{i}$ and $\tilde{%
\delta}_{i}$ symmetries, eqs.}$\left( \text{\ref{di}}\right) ,$
$\left( \text{\ref{dti}}\right) $, \textit{i.e.}
\end{itemize}
\begin{equation}
\mathcal{W}_{i}(\Sigma )\equiv\int d^{4}\!x\,\left( \phi
_{i}^{a}\frac{\delta
\Sigma }{\delta \bar{c}^{a}}+c^{a}\frac{\delta \Sigma }{\delta \bar{\phi}%
_{i}^{a}}+V_{\mu i}^{a}\frac{\delta \Sigma }{\delta \Omega _{\mu }^{a}}%
-\vartheta _{i}^{a}\frac{\delta \Sigma }{\delta L^{a}}+\frac{\delta \Sigma }{%
\delta \lambda _{i}^{a}}\frac{\delta \Sigma }{\delta b^{a}}\right) =0\;,
\label{w}
\end{equation}
\begin{equation}
\widetilde{\mathcal{W}}_{i}(\Sigma )\equiv\int d^{4}\!x\,\left( \bar{\omega}%
_{i}^{a}\frac{\delta \Sigma }{\delta \bar{c}^{a}}-c^{a}\frac{\delta \Sigma }{%
\delta \omega _{i}^{a}}-U_{\mu i}^{a}\frac{\delta \Sigma }{\delta \Omega
_{\mu }^{a}}+\eta _{i}^{a}\frac{\delta \Sigma }{\delta L^{a}}+\frac{\delta
\Sigma }{\delta \rho _{i}^{a}}\frac{\delta \Sigma }{\delta b^{a}}\right)
=0\;,  \label{wt}
\end{equation}

\begin{itemize}
\item  the $\mathcal{Q}_{i}(\Sigma )$ and $\widetilde{\mathcal{Q}}%
_{i}(\Sigma )$ identities obtained by commuting the Slavnov-Taylor
identity, eq.$\left( \text{\ref{st}}\right) ,$ with the $\mathcal{W}%
_{i}(\Sigma )$ and $\widetilde{\mathcal{W}}_{i}(\Sigma )$ identities, eqs.$%
\left( \text{\ref{w}}\right) $, $\left( \text{\ref{wt}}\right) $, namely
\begin{equation}
\mathcal{Q}_{i}(\Sigma )=0\;,  \label{qi}
\end{equation}

\begin{eqnarray}
\mathcal{Q}_{i}(\Sigma )\equiv\int d^{4}\!x\,\left( \left( \frac{\delta \Sigma }{%
\delta \lambda _{i}^{a}}+\omega _{i}^{a}\right) \frac{\delta
\Sigma }{\delta \bar{c}^{a}}+\frac{\delta \Sigma }{\delta \eta
_{i}^{a}}\frac{\delta \Sigma }{\delta b^{a}}+\left( \frac{\delta
\Sigma }{\delta \bar{\phi}_{i}^{a}}-\rho _{i}^{a}\right)
\frac{\delta \Sigma }{\delta L^{a}}+c^{a}\frac{\delta \Sigma
}{\delta \bar{\omega}_{i}^{a}}-V_{\mu i}^{a}\frac{\delta \Sigma
}{\delta \xi
_{\mu }^{a}}+N_{\mu i}^{a}\frac{\delta \Sigma }{\delta \Omega _{\mu }^{a}}%
\right) \;  \nonumber
\end{eqnarray}
\vspace{0.3cm}

\noindent and

\vspace{0.3cm}

\begin{equation}
\widetilde{\mathcal{Q}}_{i}(\Sigma )=0\;,  \label{qti}
\end{equation}
\begin{eqnarray}
\widetilde{\mathcal{Q}}_{i}(\Sigma )\equiv\int d^{4}\!x\,\left( \left( \frac{%
\delta \Sigma }{\delta \rho _{i}^{a}}-\bar{\phi}_{i}^{a}\right)
\frac{\delta
\Sigma }{\delta \bar{c}^{a}}+\frac{\delta \Sigma }{\delta \vartheta _{i}^{a}}%
\frac{\delta \Sigma }{\delta b^{a}}+\left( \frac{\delta \Sigma }{\delta
\omega _{i}^{a}}-\lambda _{i}^{a}\right) \frac{\delta \Sigma }{\delta L^{a}}%
-c^{a}\frac{\delta \Sigma }{\delta \phi _{i}^{a}}+U_{\mu
i}^{a}\frac{\delta \Sigma }{\delta \xi _{\mu }^{a}}-M_{\mu
i}^{a}\frac{\delta \Sigma }{\delta \Omega _{\mu }^{a}}\right) \;
\nonumber
\end{eqnarray}

\vspace{0.3cm}

\item  the rigid $\mathcal{R}-$identities
\begin{equation}
\mathcal{R}_{ij}^{(1)}(\Sigma )\equiv \int d^{4}\!x\,\left( \phi _{i}^{a}\frac{%
\delta \Sigma }{\delta \omega
_{j}^{a}}-\bar{\omega}_{j}^{a}\frac{\delta
\Sigma }{\delta \bar{\phi}_{i}^{a}}+V_{\mu i}^{a}\frac{\delta \Sigma }{%
\delta N_{\mu j}^{a}}+U_{\mu j}^{a}\frac{\delta \Sigma }{\delta M_{\mu i}^{a}%
}+\vartheta _{i}^{a}\frac{\delta \Sigma }{\delta \rho _{j}^{a}}-\eta _{j}^{a}%
\frac{\delta \Sigma }{\delta \lambda _{i}^{a}}\right) =0\;,  \label{r1}
\end{equation}

\begin{eqnarray}
\mathcal{R}_{ij}^{(2)}(\Sigma )\equiv\int d^{4}\!x\,\left( \phi
_{i}^{a}\frac{ \delta \Sigma }{\delta \phi
_{j}^{a}}-\bar{\phi}_{j}^{a}\frac{\delta \Sigma }{\delta
\bar{\phi}_{i}^{a}}+\omega _{i}^{a}\frac{\delta \Sigma }{\delta
\omega _{j}^{a}}-\bar{\omega}_{j}^{a}\frac{\delta \Sigma }{\delta \bar{\omega%
}_{i}^{a}}+V_{\mu i}^{a}\frac{\delta \Sigma }{\delta V_{\mu j}^{a}}-M_{\mu
j}^{a}\frac{\delta \Sigma }{\delta M_{\mu i}^{a}}\right.   \nonumber \\
\left. {\ }{\ }{\ }{\ }{\ }{\ }{\ }{\ }{\ }{\ }{\ }{\ }+N_{\mu
i}^{a}\frac{\delta \Sigma }{\delta N_{\mu j}^{a}}-U_{\mu
j}^{a}\frac{\delta \Sigma }{\delta U_{\mu i}^{a}}+\vartheta _{i}^{a}\frac{%
\delta \Sigma }{\delta \vartheta _{j}^{a}}-\eta _{j}^{a}\frac{\delta \Sigma
}{\delta \eta _{i}^{a}}+\rho _{i}^{a}\frac{\delta \Sigma }{\delta \rho
_{j}^{a}}-\lambda _{j}^{a}\frac{\delta \Sigma }{\delta \lambda _{i}^{a}}%
\right)= 0 \;,  \nonumber \\
\label{r2}
\end{eqnarray}
\begin{equation}
\mathcal{R}^{(3)}(\Sigma ) \equiv \int d^{4}\!x\,\left( \bar{\omega}_{i}^{a}\frac{%
\delta \Sigma }{\delta \omega _{i}^{a}}-U_{\mu i}^{a}\frac{\delta \Sigma }{%
\delta N_{\mu i}^{a}}-\eta _{i}^{a}\frac{\delta \Sigma }{\delta \rho _{i}^{a}%
}\right) =0\;,  \label{r3}
\end{equation}
\begin{equation}
\mathcal{R}^{(4)}(\Sigma )\equiv\int d^{4}\!x\,\left( \bar{\omega}_{i}^{a}\frac{%
\delta \Sigma }{\delta \phi
_{i}^{a}}-\bar{\phi}_{i}^{a}\frac{\delta \Sigma }{\delta \omega
_{i}^{a}}-U_{\mu i}^{a}\frac{\delta \Sigma }{\delta V_{\mu
i}^{a}}-M_{\mu i}^{a}\frac{\delta \Sigma }{\delta N_{\mu
i}^{a}}-\eta
_{i}^{a}\frac{\delta \Sigma }{\delta \vartheta _{i}^{a}}+\lambda _{i}^{a}%
\frac{\delta \Sigma }{\delta \rho _{i}^{a}}\right) =0\;,  \label{r4}
\end{equation}
\end{itemize}

The trace of $\mathcal{R}_{ij}^{(2)}$, eq.$\left( \text{\ref{r2}}\right) $, defines the $\mathcal{Q}_{f}$%
-charge of all fields and sources.

\begin{itemize}
\item  the diagonal gauge-fixing condition and the antighost equation
\end{itemize}

\begin{eqnarray}
&&\frac{\delta \Sigma }{\delta b}=\partial _{\mu }A_{\mu }\;,  \label{dgf} \\
&&  \nonumber \\
&&\frac{\delta \Sigma }{\delta \bar{c}}+\partial _{\mu }\frac{\delta \Sigma
}{\delta \Omega _{\mu }}=0\;.  \label{dagh}
\end{eqnarray}

\begin{itemize}
\item  the $\mathcal{D}(\Sigma )-$Ward identity
\end{itemize}

\begin{equation}
\mathcal{D}(\Sigma )\equiv\int d^{4}\!x\,\left( c^{a}\frac{\delta
\Sigma }{\delta \bar{c}^{a}}+\frac{\delta \Sigma }{\delta
L^{a}}\frac{\delta \Sigma }{\delta b^{a}}\right) =0\;.
\label{sl2r}
\end{equation}

\begin{itemize}
\item  the local $U(1)$ Ward identity

\begin{equation}
\mathcal{W}^{3}(\Sigma )=-\partial ^{2}b\;,  \label{u1}
\end{equation}
with
\begin{eqnarray}
\mathcal{W}^{3}\equiv\partial _{\mu }\frac{\delta }{\delta A_{\mu
}}+g\varepsilon
^{ab}\left( A_{\mu }^{a}\frac{\delta }{\delta A_{\mu }^{b}}+b^{a}\frac{%
\delta }{\delta b^{b}}+c^{a}\frac{\delta }{\delta c^{b}}+\bar{c}^{a}\frac{%
\delta }{\delta \bar{c}^{b}}+\phi _{i}^{a}\frac{\delta }{\delta \phi _{i}^{b}%
}+\bar{\phi}_{i}^{a}\frac{\delta }{\delta \bar{\phi}_{i}^{b}}\right.
\nonumber \\
+\omega _{i}^{a}\frac{\delta }{\delta \omega _{i}^{b}}+\bar{\omega}_{i}^{a}%
\frac{\delta }{\delta \bar{\omega}_{i}^{b}}+\Omega _{\mu }^{a}\frac{\delta }{%
\delta \Omega _{\mu }^{b}}+\tau _{\mu }^{a}\frac{\delta }{\delta \tau _{\mu
}^{b}}+\xi _{\mu }^{a}\frac{\delta }{\delta \xi _{\mu }^{b}}+U_{\mu i}^{a}%
\frac{\delta }{\delta U_{\mu i}^{b}}+V_{\mu i}^{a}\frac{\delta }{\delta
V_{\mu i}^{b}}  \nonumber \\
\left. +M_{\mu i}^{a}\frac{\delta }{\delta M_{\mu i}^{b}}+N_{\mu i}^{a}\frac{%
\delta }{\delta N_{\mu i}^{b}}+\eta _{i}^{a}\frac{\delta }{\delta \eta
_{i}^{b}}+\vartheta _{i}^{a}\frac{\delta }{\delta \vartheta _{i}^{b}}%
+\lambda _{i}^{a}\frac{\delta }{\delta \lambda _{i}^{b}}+\rho _{i}^{a}\frac{%
\delta }{\delta \rho _{i}^{b}}+L^{a}\frac{\delta }{\delta L^{b}}\right) \;.
\label{u1op}
\end{eqnarray}
Notice that the breaking term in the right hand side of eq.$\left(
\text{\ref {u1}}\right) $, {\it i.e.} $\partial ^{2}b$, is linear
in the quantum fields. It is thus a classical breaking, not
affected by the quantum corrections \cite{Piguet:1995er}.
\end{itemize}

\subsection{Useful commutation and anti-commutation relations}

For further use, let us give here the relevant commutation and
anti-commutation relations between the nilpotent linearized Slavnov-Taylor
operator $\mathcal{B}_{\Sigma }$,
\begin{equation}
\mathcal{B}_{\Sigma }\mathcal{B}_{\Sigma }=0\,,  \label{nilpb}
\end{equation}
\begin{eqnarray}
\mathcal{B}_{\Sigma } &=&\int d^{4}\!x\,\left( \left( \frac{\delta \Sigma }{%
\delta \Omega _{\mu }^{a}}+\frac{\delta \Sigma }{\delta \tau _{\mu }^{a}}%
\right) \frac{\delta }{\delta A_{\mu }^{a}}+\frac{\delta \Sigma }{\delta
A_{\mu }^{a}}\left( \frac{\delta }{\delta \Omega _{\mu }^{a}}+\frac{\delta }{%
\delta \tau _{\mu }^{a}}\right) +\frac{\delta \Sigma }{\delta \Omega _{\mu }}%
\frac{\delta }{\delta A_{\mu }}+\frac{\delta \Sigma }{\delta A_{\mu }}\frac{%
\delta }{\delta \Omega _{\mu }}\right.  \nonumber \\
&&+\frac{\delta \Sigma }{\delta L^{a}}\frac{\delta }{\delta c^{a}}+\frac{%
\delta \Sigma }{\delta c^{a}}\frac{\delta }{\delta L^{a}}+\frac{\delta
\Sigma }{\delta L}\frac{\delta }{\delta c}+\frac{\delta \Sigma }{\delta c}%
\frac{\delta }{\delta L}+b^{a}\frac{\delta }{\delta \bar{c}^{a}}+b\frac{%
\delta }{\delta \bar{c}}+\omega _{i}^{a}\frac{\delta }{\delta \phi _{i}^{a}}
\nonumber \\
&&\left. +\bar{\phi}_{i}^{a}\frac{\delta }{\delta \bar{\omega}_{i}^{a}}%
+N_{\mu i}^{a}\frac{\delta }{\delta V_{\mu i}^{a}}-M_{\mu i}^{a}\frac{\delta
}{\delta U_{\mu i}^{a}}-(\Omega _{\mu }^{a}-\tau _{\mu }^{a})\frac{\delta }{%
\delta \xi _{\mu }^{a}}+\lambda _{i}^{a}\frac{\delta }{\delta \eta _{i}^{a}}%
+\rho _{i}^{a}\frac{\delta }{\delta \vartheta _{i}^{a}}\right) \;,
\label{bsig}
\end{eqnarray}
and the linearized operators $\mathcal{W}_{i}^{\Sigma }$, $\widetilde{%
\mathcal{W}}_{i}^{\Sigma }$, $\mathcal{Q}_{i}^{\Sigma }$, $\widetilde{%
\mathcal{Q}}_{i}^{\Sigma }$, $\mathcal{D}_{\Sigma }$ corresponding to the
Ward identities $\left( \text{\ref{w}}\right) $, $\left( \text{\ref{wt}}%
\right) $, $\left( \text{\ref{qi}}\right) $, $\left( \text{\ref{qti}}\right)
$, $\left( \text{\ref{sl2r}}\right) $, given by
\begin{equation}
\mathcal{W}_{i}^{\Sigma }=\int d^{4}\!x\,\left( \phi _{i}^{a}\frac{\delta }{%
\delta \bar{c}^{a}}+c^{a}\frac{\delta }{\delta \bar{\phi}_{i}^{a}}+V_{\mu
i}^{a}\frac{\delta }{\delta \Omega _{\mu }^{a}}-\vartheta _{i}^{a}\frac{%
\delta }{\delta L^{a}}+\frac{\delta \Sigma }{\delta \lambda _{i}^{a}}\frac{%
\delta }{\delta b^{a}}+\frac{\delta \Sigma }{\delta b^{a}}\frac{\delta }{%
\delta \lambda _{i}^{a}}\right) \;,  \label{wl}
\end{equation}
\begin{equation}
\widetilde{\mathcal{W}}_{i}^{\Sigma }=\int d^{4}\!x\,\left( \bar{\omega}%
_{i}^{a}\frac{\delta }{\delta \bar{c}^{a}}-c^{a}\frac{\delta }{\delta \omega
_{i}^{a}}-U_{\mu i}^{a}\frac{\delta }{\delta \Omega _{\mu }^{a}}+\eta
_{i}^{a}\frac{\delta }{\delta L^{a}}+\frac{\delta \Sigma }{\delta \rho
_{i}^{a}}\frac{\delta }{\delta b^{a}}+\frac{\delta \Sigma }{\delta b^{a}}%
\frac{\delta }{\delta \rho _{i}^{a}}\right) \;,  \label{wtl}
\end{equation}

\begin{eqnarray}
\mathcal{Q}_{i}^{\Sigma } &=&\int d^{4}\!x\,\left( \left( \frac{\delta
\Sigma }{\delta \lambda _{i}^{a}}+\omega _{i}^{a}\right) \frac{\delta }{%
\delta \bar{c}^{a}}-\frac{\delta \Sigma }{\delta \bar{c}^{a}}\frac{\delta }{%
\delta \lambda _{i}^{a}}+\frac{\delta \Sigma }{\delta \eta _{i}^{a}}\frac{%
\delta }{\delta b^{a}}+\frac{\delta \Sigma }{\delta b^{a}}\frac{\delta }{%
\delta \eta _{i}^{a}}+\left( \frac{\delta \Sigma }{\delta \bar{\phi}_{i}^{a}}%
-\rho _{i}^{a}\right) \frac{\delta }{\delta L^{a}}\right.  \nonumber \\
&&\left. +\frac{\delta \Sigma }{\delta L^{a}}\frac{\delta }{\delta \bar{\phi}%
_{i}^{a}}+c^{a}\frac{\delta }{\delta \bar{\omega}_{i}^{a}}-V_{\mu i}^{a}%
\frac{\delta }{\delta \xi _{\mu }^{a}}+N_{\mu i}^{a}\frac{\delta }{\delta
\Omega _{\mu }^{a}}\right) \;,  \label{qop}
\end{eqnarray}
\begin{eqnarray}
\widetilde{\mathcal{Q}}_{i}^{\Sigma } &=&\int d^{4}\!x\left( \,\left( \frac{%
\delta \Sigma }{\delta \rho _{i}^{a}}-\bar{\phi}_{i}^{a}\right) \frac{\delta
}{\delta \bar{c}^{a}}+\frac{\delta \Sigma }{\delta \bar{c}^{a}}\frac{\delta
}{\delta \rho _{i}^{a}}+\frac{\delta \Sigma }{\delta \vartheta _{i}^{a}}%
\frac{\delta }{\delta b^{a}}+\frac{\delta \Sigma }{\delta b^{a}}\frac{\delta
}{\delta \vartheta _{i}^{a}}+\left( \frac{\delta \Sigma }{\delta \omega
_{i}^{a}}-\lambda _{i}^{a}\right) \frac{\delta }{\delta L^{a}}\right.
\nonumber \\
&&\left. +\frac{\delta \Sigma }{\delta L^{a}}\frac{\delta }{\delta \omega
_{i}^{a}}-c^{a}\frac{\delta }{\delta \phi _{i}^{a}}+U_{\mu i}^{a}\frac{%
\delta }{\delta \xi _{\mu }^{a}}-M_{\mu i}^{a}\frac{\delta }{\delta \Omega
_{\mu }^{a}}\right) \;,  \label{qopt}
\end{eqnarray}
and
\begin{equation}
\mathcal{D}_{\Sigma }=\int d^{4}\!x\,\left( c^{a}\frac{\delta }{\delta \bar{c%
}^{a}}+\frac{\delta \Sigma }{\delta L^{a}}\frac{\delta }{\delta b^{a}}+\frac{%
\delta \Sigma }{\delta b^{a}}\frac{\delta }{\delta L^{a}}\right) \;.
\label{dsig}
\end{equation}
The commutation and anti-commutation relations of these linearized operators
are found to be
\begin{eqnarray}
\{\mathcal{W}_{i}^{\Sigma },\mathcal{B}_{\Sigma }\} &=&\mathcal{Q}%
_{i}^{\Sigma }\;,  \nonumber \\
\lbrack \widetilde{\mathcal{W}}_{i}^{\Sigma },\mathcal{B}_{\Sigma }] &=&%
\widetilde{\mathcal{Q}}_{i}^{\Sigma }\;,  \nonumber \\
\lbrack \mathcal{D}_{\Sigma },\mathcal{B}_{\Sigma }] &=&0\;.  \label{rel1}
\end{eqnarray}
Furthermore, we also have
\begin{eqnarray}
\{\mathcal{R}_{ij}^{(1)},\mathcal{B}_{\Sigma }\} &=&\mathcal{R}_{ij}^{(2)}\;,
\nonumber \\
\lbrack \mathcal{R}^{(3)},\mathcal{B}_{\Sigma }] &=&\mathcal{R}^{(4)}\;,
\nonumber \\
\lbrack \frac{\delta }{\delta b},\mathcal{B}_{\Sigma }] &=&\frac{\delta }{%
\delta \bar{c}}+\partial _{\mu }\frac{\delta }{\delta \Omega _{\mu }}\;,
\nonumber \\
\lbrack \mathcal{W}^{3},\mathcal{B}_{\Sigma }] &=&0\;.  \label{rel2}
\end{eqnarray}

\section{Algebraic characterization of the most general counterterm}

\begin{table}[t]
\centering
\begin{tabular}{lcccccccc}
\hline
& $A$ & $b$ & $\bar c$ & $c$ & $\phi$ & $\bar\phi$ & $\omega$ & $\bar\omega%
\phantom{\Bigl|}$ \\ \hline\hline
dimension & $1$ & $2$ & $2$ & $0$ & $1$ & $1$ & $1$ & $1$ \\
gh number & $0$ & $0$ & $-1$ & $1$ & $0$ & $0$ & $1$ & $-1$ \\
$\mathcal{Q}_f$-charge & $0$ & $0$ & $0$ & $0$ & $1$ & $-1$ & $1$ & $-1$ \\
\hline
\end{tabular}
\caption{Quantum numbers of the fields}
\label{table2}
\end{table}

\begin{table}[t]
\centering
\begin{tabular}{lcccccccccccc}
\hline
& $\Omega$ & $\tau$ & $\xi$ & $L$ & $M$ & $N$ & $U$ & $V$ & $\eta$ & $%
\lambda $ & $\vartheta$ & $\rho$ \\ \hline\hline
dimension & $3$ & $3$ & $3$ & $4$ & $2$ & $2$ & $2$ & $2$ & $3$ & $3$ & $3$
& $3$ \\
gh number & $-1$ & $-1$ & $-2$ & $-2$ & $0$ & $1$ & $-1$ & $0$ & $-2$ & $-1$
& $-1$ & $0$ \\
$\mathcal{Q}_f$-charge & $0$ & $0$ & $0$ & $0$ & $-1$ & $1$ & $-1$ & $1$ & $%
-1$ & $-1$ & $1$ & $1$ \\ \hline
\end{tabular}
\caption{Quantum numbers of the sources}
\label{table3}
\end{table}

According to the algebraic renormalization \cite{Piguet:1995er}, the most
general local counterterm $\Sigma _{\mathrm{CT}}$ allowed by the Ward
identities $\left( \text{\ref{st}}\right) $, $\left( \text{\ref{w}}\right) $%
, $\left( \text{\ref{wt}}\right) $, $\left( \text{\ref{qi}}\right) $, $%
\left( \text{\ref{qti}}\right) $, $\left( \text{\ref{r1}}\right) $, $\left(
\text{\ref{r2}}\right) $, $\left( \text{\ref{r3}}\right) $, $\left( \text{%
\ref{r4}}\right) $, $\left( \text{\ref{dgf}}\right) $, $\left( \text{\ref
{dagh}}\right) $, $\left( \text{\ref{sl2r}}\right) $, $\left( \text{\ref{u1}}%
\right) $, is an integrated local polynomial in the fields and
sources of dimension four, see Tables $\left( \ref{table2}\right)
$ and $\left( \ref {table3}\right) $, which obeys the following
set of constraints

\begin{equation}
\mathcal{B}_{\Sigma }\,\Sigma _{\mathrm{CT}}=0\;,  \label{stl}
\end{equation}
\begin{eqnarray}
\mathcal{W}_{i}^{\Sigma }\,\Sigma _{\mathrm{CT}} &=&0\;,  \nonumber \\
\widetilde{\mathcal{W}}_{i}^{\Sigma }\,\Sigma _{\mathrm{CT}} &=&0\;,
\label{wwl}
\end{eqnarray}
\begin{eqnarray}
\mathcal{Q}_{i}^{\Sigma }\,\Sigma _{\mathrm{CT}} &=&0\;,  \nonumber \\
\widetilde{\mathcal{Q}}_{i}^{\Sigma }\,\Sigma _{\mathrm{CT}} &=&0\;,
\label{qql}
\end{eqnarray}
\begin{eqnarray}
\mathcal{R}_{ij}^{(1)}\,\Sigma _{\mathrm{CT}} &=&0\;,  \nonumber \\
\mathcal{R}_{ij}^{(2)}\,\Sigma _{\mathrm{CT}} &=&0\;,  \nonumber \\
\mathcal{R}^{(3)}\,\Sigma _{\mathrm{CT}} &=&0\;,  \nonumber \\
\mathcal{R}^{(4)}\,\Sigma _{\mathrm{CT}} &=&0\;,  \label{rr}
\end{eqnarray}
\begin{equation}
\mathcal{D}_{\Sigma }\,\Sigma _{\mathrm{CT}}=0\;,  \label{d}
\end{equation}
\begin{equation}
\mathcal{W}^{3}\,\Sigma _{\mathrm{CT}}=0\;,  \label{u1w}
\end{equation}
\begin{eqnarray}
\frac{\delta \Sigma _{\mathrm{CT}}}{\delta b}\, &=&0\;,  \nonumber \\
\frac{\delta \Sigma _{\mathrm{CT}}}{\delta \bar{c}}+\partial _{\mu }\frac{%
\delta \Sigma _{\mathrm{CT}}}{\delta \Omega _{\mu }}\, &=&0\;.  \label{gf}
\end{eqnarray}
From eqs.$\left( \text{\ref{gf}}\right) $ it follows that $\Sigma _{\mathrm{%
CT}}$ is independent of the diagonal Lagrange multiplier $b$, and
that the diagonal antighost $\bar{c}$ enters only through the
combination $(\Omega _{\mu }+\partial _{\mu }\bar{c})$.
Furthermore, from general results on the cohomology of gauge
theories \cite{Piguet:1995er,Barnich:2000zw}, it turns
out that the most general solution of the constraint $\left( \text{\ref{stl}}%
\right) $ can be written as
\begin{equation}
\Sigma _{\mathrm{CT}}=a_{0}\,S_{\mathrm{YM}}+\mathcal{B}_{\Sigma }\Delta
^{(-1)}\;,  \label{stlsol}
\end{equation}
with $\Delta ^{(-1)}$ being an integrated local polynomial with ghost number
-1, given by

\begin{eqnarray}
\Delta ^{(-1)} &=&\int d^{4}\!x\,\left( a_{1}\,\Omega _{\mu }^{a}A_{\mu
}^{a}+a_{2}\,\tau _{\mu }^{a}A_{\mu }^{a}+a_{3}\,\xi _{\mu
}^{a}\,g\varepsilon ^{ab\!}A_{\mu }^{b}c+a_{4}\,\xi _{\mu }^{a}\,\partial
_{\mu }c^{a}+a_{5}\,\xi _{\mu }^{a}\,g\varepsilon ^{ab\!}A_{\mu
}c^{b}+a_{6}\,(\partial _{\mu }\bar{c}^{a})A_{\mu }^{a}\right.  \nonumber \\
&&+a_{7}\,(\Omega _{\mu }+\partial _{\mu }\bar{c})A_{\mu
}+a_{8}\,c^{a}L^{a}+a_{9}\,cL+a_{10}\,\eta _{i}^{a}\,g\varepsilon
^{ab\!}\phi _{i}^{b}c+a_{11}\,\eta _{i}^{a}\omega _{i}^{a}+a_{12}\,\vartheta
_{i}^{a}\,g\varepsilon ^{ab\!}\bar{\omega}_{i}^{b}c  \nonumber \\
&&+a_{13}\,\vartheta _{i}^{a}\bar{\phi}_{i}^{a}+a_{14}\,\lambda _{i}^{a}\phi
_{i}^{a}+a_{15}\,\rho _{i}^{a}\bar{\omega}_{i}^{a}+a_{16}\,U_{\mu
i}^{a}\,\partial _{\mu }\phi _{i}^{a}+a_{17}\,U_{\mu i}^{a}\,g\varepsilon
^{ab\!}A_{\mu }\phi _{i}^{b}+a_{18}V_{\mu i}^{a}\,\partial _{\mu }\bar{\omega%
}_{i}^{a}  \nonumber \\
&&+a_{19}\,V_{\mu i}^{a}\,g\varepsilon ^{ab\!}A_{\mu }\bar{\omega}%
_{i}^{b}+a_{20}\,\bar{c}^{a}b^{a}+a_{21}\,g\varepsilon ^{ab\!}\bar{c}^{a}%
\bar{c}^{b}c+a_{22}\,g\varepsilon ^{ab\!}\bar{c}^{a}A_{\mu }A_{\mu
}^{b}+a_{23}\,\bar{\omega}_{i}^{a}\phi _{i}^{a}\bar{\phi}_{j}^{b}\phi
_{j}^{b}  \nonumber \\
&&+a_{24}\,\bar{\omega}_{i}^{a}\phi _{i}^{a}\bar{\omega}_{j}^{b}\omega
_{j}^{b}+a_{25}\,\bar{\omega}_{i}^{a}\phi _{i}^{b}\bar{\phi}_{j}^{a}\phi
_{j}^{b}+a_{26}\,\bar{\omega}_{i}^{a}\phi _{i}^{b}\bar{\omega}_{j}^{a}\omega
_{j}^{b}+a_{27}\,\bar{\omega}_{i}^{a}\phi _{i}^{b}\bar{\phi}_{j}^{b}\phi
_{j}^{a}+a_{28}\,\bar{\omega}_{i}^{a}\phi _{i}^{b}\bar{\omega}_{j}^{b}\omega
_{j}^{a}  \nonumber \\
&&+a_{29}\,\bar{\omega}_{i}^{a}\phi _{j}^{a}\bar{\phi}_{i}^{b}\phi
_{j}^{b}+a_{30}\,\bar{\omega}_{i}^{a}\phi _{j}^{b}\bar{\phi}_{i}^{a}\phi
_{j}^{b}+a_{31}\,\bar{\omega}_{i}^{a}\phi _{j}^{a}\bar{\omega}_{i}^{b}\omega
_{j}^{b}+a_{32}\,\bar{\omega}_{i}^{a}\phi _{i}^{a}\bar{c}^{b}c^{b}+a_{33}\,%
\bar{\omega}_{i}^{a}\phi _{i}^{b}\bar{c}^{a}c^{b}  \nonumber \\
&&+a_{34}\,\bar{\omega}_{i}^{a}\phi _{i}^{b}\bar{c}^{b}c^{a}+a_{35}\,\bar{%
\omega}_{i}^{a}\phi _{i}^{a}A_{\mu }A_{\mu }+a_{36}\,\bar{\omega}%
_{i}^{a}\phi _{i}^{a}A_{\mu }^{b}A_{\mu }^{b}+a_{37}\,\bar{\omega}%
_{i}^{a}\phi _{i}^{b}A_{\mu }^{a}A_{\mu }^{b}+a_{38}\,\bar{\omega}%
_{i}^{a}\partial ^{2}\phi _{i}^{a}  \nonumber \\
&&\left. +a_{39}\,\bar{\omega}_{i}^{a}\,g\varepsilon ^{ab\!}A_{\mu }\partial
_{\mu }\phi _{i}^{b}+a_{40}\,\bar{\omega}_{i}^{a}\,g\varepsilon
^{ab\!}(\partial _{\mu }A_{\mu })\phi _{i}^{b}+a_{41}\,\chi U_{\mu
i}^{a}V_{\mu i}^{a}\right) \;,  \nonumber \\
&&  \label{count}
\end{eqnarray}
where the coefficients $a_{i}$, $i=0,...,41$ are free dimensionless
parameters. Notice also that in the derivation of expression $\left( \text{%
\ref{count}}\right) $ use has been made of the fact that the action $\Sigma $%
, and thus $\Sigma _{\mathrm{CT}}$, are left invariant by the following
discrete symmetry
\begin{eqnarray}
\Psi ^{1} &\to &\Psi ^{1}\;,  \nonumber \\
\Psi ^{2} &\to &-\Psi ^{2}\;,  \nonumber \\
\Psi ^{\mathrm{diag}} &\to &-\Psi ^{\mathrm{diag}}\;,
\label{chage-conjugation}
\end{eqnarray}
where $\Psi ^{a}$, $a=1,2$, and $\Psi ^{\mathrm{diag}}$ stand,
respectively, for all off-diagonal and diagonal fields and
external sources. As one can easily recognize, this symmetry plays
the role of the charge conjugation \cite{Fazio:2001rm}. \newline
\newline
After a rather long calculation, from the constraints $\left( \text{\ref{wwl}%
}\right) $, $\left( \text{\ref{qql}}\right) $, $\left( \text{\ref{rr}}%
\right) $, $\left( \text{\ref{d}}\right) $, $\left( \text{\ref{u1w}}\right) $%
, one finds that
\begin{eqnarray}
a_{7} &=&a_{10}=a_{11}=a_{12}=a_{13}=a_{14}=a_{15}=0\;,  \nonumber \\
a_{25} &=&a_{26}=a_{27}=a_{28}=a_{29}=a_{30}=a_{31}=0\;,  \nonumber \\
a_{33} &=&a_{34}=0\;,  \label{van}
\end{eqnarray}
while

\begin{eqnarray}
a_{3} &\!\!\!=\!\!\!&a_{1}\;,  \nonumber \\
a_{5} &\!\!\!=\!\!\!&-a_{4}\;,  \nonumber \\
a_{22} &\!\!\!=\!\!\!&a_{6}\;,  \nonumber \\
a_{16} &\!\!\!=\!\!\!&-a_{17}=-a_{18}=a_{19}=a_{1}-a_{4}+a_{8}\;,  \nonumber
\\
a_{21} &=&-a_{20}\;,  \nonumber \\
a_{23} &=&-a_{24}=g^{2}a_{20}-\alpha g^{2}a_{8}\;,  \nonumber \\
a_{32} &=&-2g^{2}a_{20}+\alpha g^{2}a_{8}\;,  \nonumber \\
a_{38} &\!\!\!=\!\!\!&-a_{40}=-\frac{a_{39}}{2}=-\frac{a_{35}}{g^{2}}=\frac{%
a_{36}}{g^{2}}=-\frac{a_{37}}{g^{2}}=a_{6}+a_{8}\;.  \label{relc}
\end{eqnarray}
Making use of the relation
\begin{equation}
\tau _{\mu }^{a}A_{\mu }^{a}=\Omega _{\mu }^{a}A_{\mu }^{a}+\xi _{\mu
}^{a}(D_{\mu }^{ab}c^{b}+g\varepsilon ^{ab}A_{\mu }^{b}c)+\mathcal{B}%
_{\Sigma }(\xi _{\mu }^{a}A_{\mu }^{a})\;,  \label{rt}
\end{equation}
and remembering that $\mathcal{B}_{\Sigma }^{2}=0$, we obtain
\begin{eqnarray}
\Delta ^{(-1)} &=&\int d^{4}\!x\left( \,(a_{1}+a_{2})[\Omega _{\mu
}^{a}A_{\mu }^{a}+g\varepsilon ^{ab}\xi _{\mu }^{a}A_{\mu
}^{b}c]+(a_{2}+a_{4})\xi _{\mu }^{a}D_{\mu }^{ab}c^{b}\right.   \nonumber \\
&&+(a_{1}-a_{4}+a_{8})[U_{\mu i}^{a}D_{\mu }^{ab}\phi _{i}^{b}-V_{\mu
i}^{a}D_{\mu }^{ab}\bar{\omega}_{i}^{a}]-(a_{6}+a_{8})\,\bar{\omega}_{i}^{a}%
\mathcal{M}^{ab}\phi _{i}^{a}  \nonumber \\
&&-a_{6}\,\bar{c}^{a}D_{\mu }^{ab}A_{\mu
}^{b}+a_{8}\,L^{a}c^{a}+a_{9}\,Lc+a_{41}\,\chi U_{\mu i}^{a}V_{\mu i}^{a}\;%
\text{ }+a_{20}\left( \bar{c}^{a}b^{a}-g\varepsilon ^{ab}\bar{c}^{a}\bar{c}%
^{b}c\right)   \nonumber \\
&&\left. -\left( \alpha a_{8}-a_{20}\right) g^{2}\bar{\omega}_{i}^{a}\phi
_{i}^{a}\left( \bar{\phi}_{j}^{b}\phi _{j}^{b}-\bar{\omega}_{j}^{b}\omega
_{j}^{b}\right) +\left( \alpha a_{8}-2a_{20}\right) g^{2}\bar{\omega}%
_{i}^{a}\phi _{i}^{a}\overline{c}^{b}c^{b}\right)   \label{ex}
\end{eqnarray}
Finally, by renaming the coefficients as follows
\begin{eqnarray}
a_{1}+a_{2} &\to &a_{1}\;,  \nonumber \\
a_{2}+a_{4} &\to &-a_{2}\;,  \nonumber \\
a_{8} &\to &a_{3}\;,  \nonumber \\
a_{6} &\to &a_{4}\;,  \nonumber \\
a_{9} &\to &a_{5}\;,  \nonumber \\
a_{41} &\to &a_{6}\;,  \nonumber \\
a_{20} &\rightarrow &\frac{\alpha }{2}a_{7}  \label{rn}
\end{eqnarray}
for the most general local allowed counterterm $\Sigma _{\mathrm{CT}}$, we
obtain
\begin{eqnarray}
\Sigma _{\mathrm{CT}} &=&\int d^{4}\!x\,\left( (a_{0}+2a_{1})\biggl[\frac{1}{%
2}(\partial _{\mu }A_{\nu }^{a})(\partial _{\mu }A_{\nu }^{a}-\partial _{\nu
}A_{\mu }^{a})-g\varepsilon ^{ab}(\partial _{\mu }A_{\nu }^{a})(A_{\mu
}A_{\nu }^{b}-A_{\nu }A_{\mu }^{b})\right.   \nonumber \\
&&+g\varepsilon ^{ab}(\partial _{\mu }A_{\nu }^{a})A_{\mu }^{a}A_{\mu }^{b}+%
\frac{g^{2}}{2}(A_{\mu }A_{\mu }A_{\nu }^{a}A_{\nu }^{a}+A_{\mu }A_{\nu
}A_{\mu }^{a}A_{\nu }^{a})\biggr]+\frac{a_{0}}{2}\,(\partial _{\mu }A_{\nu
})(\partial _{\mu }A_{\nu }-\partial _{\nu }A_{\mu })  \nonumber \\
&&+(a_{0}+4a_{1})\frac{g^{2}}{4}A_{\mu }^{a}A_{\mu }^{a}A_{\nu }^{b}A_{\nu
}^{b}+(a_{1}-a_{4})b^{a}D_{\mu }^{ab}A_{\mu }^{b}  \nonumber \\
&&-(a_{3}+a_{4})[\bar{c}^{a}\partial ^{2}c^{a}-\bar{c}^{a}g\varepsilon
^{ab}(\partial _{\mu }A_{\mu })c^{b}-2\bar{c}^{a}g\varepsilon ^{ab}A_{\mu
}\partial _{\mu }c^{b}-g^{2}\bar{c}^{a}c^{a}A_{\mu }A_{\mu }]  \nonumber \\
&&+(a_{3}+a_{4})[\bar{\phi}_{i}^{a}\partial ^{2}\phi _{i}^{a}-\bar{\phi}%
_{i}^{a}g\varepsilon ^{ab}(\partial _{\mu }A_{\mu })\phi _{i}^{b}-2\bar{\phi}%
_{i}^{a}g\varepsilon ^{ab}A_{\mu }\partial _{\mu }\phi _{i}^{b}-g^{2}\bar{%
\phi}_{i}^{a}\phi _{i}^{a}A_{\mu }A_{\mu }]  \nonumber \\
&&-(a_{3}+a_{4})[\bar{\omega}_{i}^{a}\partial ^{2}\omega _{i}^{a}-\bar{\omega%
}_{i}^{a}g\varepsilon ^{ab}(\partial _{\mu }A_{\mu })\omega _{i}^{b}-2\bar{%
\omega}_{i}^{a}g\varepsilon ^{ab}A_{\mu }\partial _{\mu }\omega
_{i}^{b}-g^{2}\bar{\omega}_{i}^{a}\omega _{i}^{a}A_{\mu }A_{\mu }]  \nonumber
\\
&&+(2a_{1}-a_{3}-a_{4})g^{2}\varepsilon ^{ac}\varepsilon ^{bd}(\bar{c}%
^{a}c^{b}-\bar{\phi}_{i}^{a}\phi _{i}^{b}+\bar{\omega}_{i}^{a}\omega
_{i}^{b})A_{\mu }^{c}A_{\mu }^{d}  \nonumber \\
&&+(a_{1}-a_{4}-a_{5})g\varepsilon ^{ab}\bar{c}^{a}cD_{\mu }^{bc}A_{\mu
}^{c}-(a_{1}-a_{3})(\Omega _{\mu }+\partial _{\mu }\bar{c})g\varepsilon
^{ab}A_{\mu }^{a}c^{b}  \nonumber \\
&&+(a_{1}-2a_{3}-a_{4})[2g^{2}\varepsilon ^{ab}\varepsilon ^{cd}\bar{\omega}%
_{i}^{a}A_{\mu }^{c}c^{d}\partial_{\mu }\phi
_{i}^{b}+g^{2}\varepsilon ^{ab}\varepsilon
^{cd}\bar{\omega}_{i}^{a}\partial_{\mu}(A_{\mu }^{c}c^{d})\phi
_{i}^{b}  \nonumber \\
&&+2g^{3}\varepsilon ^{bc}\bar{\omega}_{i}^{a}\phi _{i}^{a}A_{\mu
}A_{\mu }^{b}c^{c}-g^{2}(\varepsilon ^{ac}\varepsilon
^{bd}+\varepsilon ^{ad}\varepsilon
^{bc})\bar{\omega}_{i}^{a}A_{\mu }^{d}(D_{\mu
}^{ce}c^{e})\phi _{i}^{b}]  \nonumber \\
&&+(2a_{1}-a_{3}-a_{4}-a_{5})g^{3}(\delta ^{ae}\varepsilon ^{bd}+\delta
^{be}\varepsilon ^{ad})\,\bar{\omega}_{i}^{a}A_{\mu }^{d}A_{\mu }^{e}\phi
_{i}^{b}c  \nonumber \\
&&-(a_{2}+2a_{3})g^{2}\varepsilon ^{ab}\varepsilon ^{cd}(U_{\mu i}^{a}\phi
_{i}^{b}+V_{\mu i}^{a}\bar{\omega}_{i}^{b})A_{\mu
}^{c}c^{d}+(a_{1}+a_{2}+a_{3})\Omega _{\mu }^{a}D_{\mu }^{ab}c^{b}  \nonumber
\\
&&-a_{2}\tau _{\mu }^{a}D_{\mu }^{ab}c^{b}+a_{5}g\varepsilon ^{ab}\tau _{\mu
}^{a}A_{\mu }^{b}c-(a_{1}+a_{2}+a_{3}+a_{5})g\varepsilon ^{ab}\xi _{\mu
}^{a}(D_{\mu }^{ab}c^{c})c  \nonumber \\
&&+(a_{2}+2a_{3})\frac{g^{2}}{2}\varepsilon ^{ab}\varepsilon ^{cd}\xi _{\mu
}^{a}A_{\mu }^{b}c^{c}c^{d}-a_{5}g\varepsilon ^{ab}L^{a}c^{b}c-(2a_{3}-a_{5})%
\frac{g^{2}}{2}\varepsilon ^{ab}Lc^{a}c^{b}  \nonumber \\
&&+a_{5}(\Omega _{\mu }+\partial _{\mu }\bar{c})\partial _{\mu
}c-a_{3}g^{2}\varepsilon ^{ab}\varepsilon ^{cd}(\eta _{i}^{a}\phi
_{i}^{b}+\vartheta _{i}^{a}\bar{\omega}_{i}^{b})c^{c}c^{d}-a_{5}g\varepsilon
^{ab}(\lambda _{i}^{a}\phi _{i}^{b}+\eta _{i}^{a}\omega _{i}^{b}  \nonumber
\\
&&+\rho _{i}^{a}\bar{\omega}_{i}^{b}-\vartheta _{i}^{a}\bar{\phi}%
_{i}^{b})c-(a_{3}+a_{4}+a_{5})[2g\varepsilon ^{ab}\bar{\omega}%
_{i}^{a}(\partial _{\mu }c)\partial _{\mu }\phi _{i}^{b}+g\varepsilon ^{ab}%
\bar{\omega}_{i}^{a}(\partial ^{2}c)\phi _{i}^{b}  \nonumber \\
&&+2g^{2}\bar{\omega}_{i}^{a}\phi _{i}^{a}A_{\mu }\partial _{\mu
}c]+(a_{1}+a_{2}+a_{3}+a_{5})g\varepsilon ^{ab}(\partial _{\mu }c)(U_{\mu
i}^{a}\phi _{i}^{b}-V_{\mu i}^{a}\bar{\omega}_{i}^{b})  \nonumber \\
&&\left. -(a_{1}+a_{2}+a_{3})(M_{\mu i}^{a}D_{\mu }^{ab}\phi _{i}^{b}+U_{\mu
i}^{a}D_{\mu }^{ab}\omega _{i}^{b}+V_{\mu i}^{a}D_{\mu }^{ab}\bar{\phi}%
_{i}^{b}+N_{\mu i}^{a}D_{\mu }^{ab}\bar{\omega}_{i}^{b}) \right. \nonumber \\
&&\left. -a_{6}\,\chi (M_{\mu i}^{a}V_{\mu i}^{a}+U_{\mu i}^{a}N_{\mu i}^{a})\right)   \nonumber \\
&&+\frac{\alpha }{2}\int d^{4}x\left( a_{7}b^{a}b^{a}-2\left(
a_{7}-a_{5}\right) g\varepsilon ^{ab}b^{a}\overline{c}^{b}c+g^{2}\left(
a_{7}-2a_{3}\right) \overline{c}^{a}c^{a}\overline{c}^{b}c^{b}\right.
\nonumber \\
&& {\ }{\ }{\ }{\ }{\ }{\ }+g^{2}\left( a_{7}-2a_{3}\right) \left( \bar{\phi}_{i}^{a}\phi _{i}^{a}-%
\bar{\omega}_{i}^{a}\omega _{i}^{a}\right) \left( \bar{\phi}_{j}^{b}\phi
_{j}^{b}-\bar{\omega}_{j}^{b}\omega _{j}^{b}\right) -2g^{2}\left(
a_{7}-2a_{3}\right) \left( \bar{\phi}_{i}^{a}\phi _{i}^{a}-\bar{\omega}%
_{i}^{a}\omega _{i}^{a}\right) \overline{c}^{b}c^{b}  \nonumber \\
&&\left. {\ }{\ }{\ }{\ }{\ }{\ } +2g^{2}\left(
a_{7}-2a_{3}\right) \bar{\omega}_{i}^{a}\phi
_{i}^{a}b^{b}c^{b}-2g^{3}\left( a_{7}-2a_{3}-a_{5}\right) \bar{\omega}%
_{i}^{a}\phi _{i}^{a}\varepsilon ^{bc}\overline{c}^{b}c^{c}c\right) \;.
\label{CT}
\end{eqnarray}

\subsection{Stability of the classical action and renormalization factors}

After the characterization of the most general local counterterm $\Sigma _{%
\mathrm{CT}}$, eq.$\left( \text{\ref{CT}}\right) $, compatible with all Ward
identities, we still have to check that it can be reabsorbed through a
multiplicative redefinition of the fields, sources and parameters of the
starting action $\Sigma $, according to

\begin{equation}
\Sigma (\Psi _{0},\psi _{0},J_{0},\Omega _{0},\tau _{0},\zeta _{0})=\Sigma
(\Psi ,\psi ,J,\Omega ,\tau ,\zeta )+\epsilon \Sigma _{\mathrm{CT}}(\Psi
,\psi ,J,\Omega ,\tau ,\zeta )\;+\;O(\epsilon ^{2})\;,  \label{rb1}
\end{equation}
\noindent where
\begin{equation}
\Psi _{0}={\widetilde{Z}}_{\Psi }^{1/2}\,\Psi \;,  \label{Psi}
\end{equation}
denotes the off-diagonal fields, $\Psi \equiv \left( A_{\mu }^{a},b^{a},c^{a},%
\bar{c}^{a}\right) $, while
\begin{equation}
\psi _{0}=Z_{\psi }^{1/2}\,\psi \;,  \label{psi}
\end{equation}
stands for the fields, $\psi \equiv \left( A_{\mu },b,c,\bar{c},\phi
_{i}^{a},\bar{\phi}_{i}^{a},\omega _{i}^{a},\bar{\omega}_{i}^{a}\right) $.
\newline
\newline
Also $J$ and $\zeta $
\begin{eqnarray}
J_{0} &=&Z_{J}\,J\;,  \nonumber \\
\zeta _{0} &=&Z_{\zeta }\,\zeta \;,  \label{jxi}
\end{eqnarray}
denote the external sources, $J\equiv \left( \xi _{\mu }^{a},L^{a},L,\Omega
_{\mu },U_{\mu i}^{a},V_{\mu i}^{a},M_{\mu i}^{a},N_{\mu i}^{a},\eta
_{i}^{a},\lambda _{i}^{a},\vartheta _{i}^{a},\rho _{i}^{a}\right) $, and the
parameters $\zeta \equiv $ $\left( g,\chi ,\alpha \right) $, respectively.
\newline
\newline
Moreover, recalling that the sources $\Omega _{\mu }^{a}$ and $\tau _{\mu
}^{a}$ are coupled to composite operators, $\mathcal{O}_{1}=-D_{\mu
}^{ab}c^{b},$ $\mathcal{O}_{2}=-g\varepsilon ^{ab}A_{\mu }^{b}c$, displaying
the same quantum numbers, see eqs.$\left( \text{\ref{2opd}}\right) $,$\left(
\text{\ref{2opsou}}\right) $, they renormalize as
\begin{equation}
\begin{pmatrix} \Omega_{0\mu}^{\phantom{0}a}\cr \tau_{0\mu}^{\phantom{0}a}
\end{pmatrix} =Z_{\Omega \tau }\begin{pmatrix} \Omega_{\mu}^{a}\cr
\tau_{\mu}^{a} \end{pmatrix}\;,  \label{rm1}
\end{equation}
where the matrix
\begin{equation}
Z_{\Omega \tau }=1+\epsilon \begin{pmatrix} z_{\Omega}&z_{\Omega\tau}\cr
0&z_{\tau} \end{pmatrix}\;,  \label{rm2}
\end{equation}
allows for the mixing at the quantum level between the operators $\mathcal{O}%
_{1},\mathcal{O}_{2}$. By direct inspection of $\Sigma _{\mathrm{CT}}$, the
renormalization factors are easily found to be
\begin{eqnarray}
{\widetilde{Z}}_{b} &=&Z_{g}^{2}Z_{\bar{c}}^{-1}{\widetilde{Z}}_{c}\;,
\nonumber \\
Z_{b} &=&Z_{g}^{2}\;,  \nonumber \\
Z_{\Omega } &=&Z_{\bar{c}}^{1/2}\;,  \nonumber \\
Z_{\bar{\omega}} &=&Z_{g}^{-2}Z_{\bar{c}}{\widetilde{Z}}_{c}\;,  \nonumber \\
Z_{\omega } &=&Z_{g}^{2}Z_{\bar{c}}^{-1}{\widetilde{Z}}_{c}\;,  \nonumber \\
Z_{\phi } &=&\;Z_{\bar{\phi}}\;=\;{\widetilde{Z}}_{c}\;,  \nonumber \\
Z_{U} &=&Z_{g}^{-1}Z_{\bar{c}}^{1/2}{Z}_{V}\;,  \nonumber \\
Z_{N} &=&Z_{g}Z_{\bar{c}}^{-1/2}{Z}_{V}\;,\;  \nonumber \\
Z_{M} &=&Z_{V}\;,\;\;  \nonumber \\
Z_{\eta } &=&Z_{g}^{-2}Z_{\bar{c}}{\widetilde{Z}}_{c}^{-1/2}\;,  \nonumber \\
Z_{\lambda } &=&Z_{\vartheta }\;\;=\;\;Z_{g}^{-1}Z_{\bar{c}}^{1/2}{%
\widetilde{Z}}_{c}^{-1/2}\;,  \nonumber \\
Z_{\rho }\;\; &=&\;\;{\widetilde{Z}}_{c}^{-1/2}\;,  \nonumber \\
Z_{\xi } &=&Z_{V}\;,  \nonumber \\
Z_{L^{a}} &=&Z_{g}^{-1}Z_{\bar{c}}^{1/2}{\widetilde{Z}}_{c}^{-1/2}\;,
\nonumber \\
Z_{L} &=&Z_{g}^{-1}Z_{\bar{c}}^{1/2}Z_{c}^{-1/2}\;,  \nonumber \\
Z_{A} &=&Z_{g}^{-2}\;,  \label{zf}
\end{eqnarray}
with

\begin{eqnarray}
{\widetilde{Z}}_{A} &=&1+\epsilon \,(a_{0}+2a_{1})\;,  \nonumber \\
Z_{g} &=&1-\epsilon \,\frac{a_{0}}{2}\;,  \nonumber \\
{\widetilde{Z}_{c}} &=&{\widetilde{Z}_{\bar{c}}}\;\;=\;\;1-\epsilon
\,(a_{3}+a_{4})\;,  \nonumber \\
Z_{c} &=&1+\epsilon \,(a_{3}-a_{4}-2a_{5})\;,  \nonumber \\
Z_{\bar{c}} &=&1-\epsilon \,(a_{3}-a_{4})\;,  \nonumber \\
Z_{V}\;\; &=&\;\;1+\epsilon \,\left( -a_{1}-a_{2}-\frac{a_{3}}{2}+\frac{a_{4}%
}{2}\right) \;,  \nonumber \\
Z_{\chi } &=&1+\epsilon \,(2a_{1}+2a_{2}+a_{3}-a_{4}-a_{6})\;\;,  \nonumber
\\
Z_{\alpha } &=&1+\epsilon \,(a_{7}+a_{0}+2a_{4})\;\;.  \label{zf2}
\end{eqnarray}
Finally
\begin{equation}
Z_{\Omega \tau }=\begin{pmatrix} Z_{V}&\/& Z_{g}^{-1}Z_{\bar
c}^{1/2}Z_{V}^{-1}{\widetilde Z}_{A}^{-1/2}-1\cr \/&\/&\cr 0
&\/&Z_{g}^{-1}Z_{\bar c}^{1/2}{\widetilde Z}_{A}^{-1/2} \end{pmatrix}\;.
\label{zfm}
\end{equation}
Equations $\left( \text{\ref{zf}}\right) ,\left( \text{\ref{zf2}}\right)
,\left( \text{\ref{zfm}}\right) $ show that the counterterm $\Sigma _{%
\mathrm{CT}}$ can be reabsorbed by means of a redefinition of the fields,
sources and parameters of the starting action $\Sigma $, establishing thus
the multiplicative renormalizability of the theory. Let us end this section
by noting that the renormalization factors $\left( Z_{\phi },Z_{\bar{\phi}%
},Z_{\bar{\omega}},Z_{\omega }\right) $ of the auxiliary localizing fields $%
\left( \phi _{i}^{a},\bar{\phi}_{i}^{a},\omega _{i}^{a},\bar{\omega}%
_{i}^{a}\right) \;$are not independent quantities, being expressed in terms
of the renormalization factors of the gauge coupling constant and of the
Faddeev-Popov ghosts. Again, this feature stems from the Ward identities $%
\left( \text{\ref{w}}\right) $, $\left( \text{\ref{wt}}\right) $. \newline
\newline
We also notice that the renormalization factor $Z_{g}$ of the gauge coupling
constant $g$ can be obtained directly from the renormalization factor $Z_{A}$
of the diagonal component of the gauge field $A_{\mu }$. Thus, the
nonrenormalization theorem of the maximal Abelian gauge \cite
{Fazio:2001rm,Dudal:2004rx}
\begin{equation}
Z_{A}Z_{g}^{2}=1\;,  \label{nren}
\end{equation}
remains valid in the presence of the Gribov horizon. This important property
is a direct consequence of the local $U(1)$ Ward identity $\left( \ref{u1}%
\right) $.

\section{Introduction of a generalized dimension two local operator}

As remarked in the Introduction, the inclusion of the horizon term, eq.$%
\left( \ref{i8}\right) $, does not prevent us from defining a local
composite dimension two operator $\mathcal{O}_{A^{2}}$%
\begin{equation}
\mathcal{O}_{A^{2}}\equiv \frac{1}{2}A_{\mu }^{a}A_{\mu }^{a}+\alpha \left(
\overline{c}^{a}c^{a}-\bar{\phi}_{i}^{a}\phi _{i}^{a}+\bar{\omega}%
_{i}^{a}\omega _{i}^{a}\right) \;,  \label{oa2}
\end{equation}
which generalizes the gluon operator $\left( \frac{1}{2}A_{\mu }^{a}A_{\mu
}^{a}+\alpha \overline{c}^{a}c^{a}\right) $ already considered in the
maximal Abelian gauge \cite{Kondo:2001tm}, and proven to be renormalizable
to all orders \cite{Dudal:2004rx}. From the equation of motion of the
Lagrange multiplier $b^{a}$

\begin{equation}
\frac{\delta \Sigma }{\delta b^{a}}=D_{\mu }^{ab}A_{\mu }^{b}+\alpha \left(
b^{a}-g\varepsilon ^{ab}\overline{c}^{b}c+g^{2}\bar{\omega}_{i}^{b}\phi
_{i}^{b}c^{a}\right) \;,  \label{a1}
\end{equation}
it follows that the integrated operator $\int d^{4}x\mathcal{O}_{A^{2}}$
enjoys the property of being $BRST\;$invariant on-shell, namely
\begin{equation}
s\int d^{4}x\mathcal{O}_{A^{2}}=\int d^{4}x\left( -\left( D_{\mu
}^{ab}c^{b}\right) A_{\mu }^{a}+\alpha b^{a}c^{a}-\alpha g\overline{c}%
^{a}\varepsilon ^{ab}c^{b}c\right) =\int d^{4}xc^{a}\frac{\delta \Sigma }{%
\delta b^{a}}\;.  \label{a3}
\end{equation}
\newline
As proven in the Landau gauge
\cite{Verschelde:2001ia,Dudal:2002pq,Dudal:2005na} in the case of
the operator $A_{\mu }^{A}A_{\mu }^{A}$, $A=1,...,N^{2}-1$,
eq.$\left( \ref{a3}\right) $ implies that the local composite
operator $\mathcal{O}_{A^{2}}$ is
multiplicatively renormalizable. This can be established by introducing $%
\mathcal{O}_{A^{2}}$ in the starting action $\Sigma $, and by
repeating the algebraic analysis done in the previous section.
More precisely, we introduce the operator $\mathcal{O}_{A^{2}}$ by
means of a $BRST$ doublet of external sources $(J,\Lambda )$
\begin{equation}
s\Lambda =J,\qquad sJ=0\;,  \label{a4}
\end{equation}
so that
\begin{eqnarray}
\widetilde{\Sigma } &=&\Sigma +s\int d^{4}\!x\,\left( \Lambda \mathcal{O}%
_{A^{2}}+\frac{\varrho }{2}\Lambda J\right)  \nonumber \\
&=&\Sigma +\int d^{4}\!x\left( \,J\mathcal{O}_{A^{2}}+\Lambda
\,\left(
A_{\mu }^{a}D_{\mu }^{ab}c^{b}-\alpha b^{a}c^{a}+\alpha g\overline{c}%
^{a}\varepsilon ^{ab}c^{b}c\right) +\frac{\varrho
}{2}J^{2}\right) \;. \label{a5}
\end{eqnarray}
The parameter $\varrho $ is a free parameter, needed in order to
account for the ultraviolet divergences  of the correlation
function $\left\langle \left( A_{\mu }^{a}(x)A_{\mu
}^{a}(x)\right) \left( A_{\nu }^{b}(y)A_{\nu }^{b}(y)\right)
\right\rangle $ \cite{Dudal:2004rx}. \newline
\newline
The extended action $\widetilde{\Sigma }$ is easily seen to obey the
following Slavnov-Taylor identity
\begin{equation}
\mathcal{S}(\widetilde{\Sigma })=0\;,  \label{a6}
\end{equation}
with{\
\begin{eqnarray}
\mathcal{S}(\widetilde{\Sigma }) &=&\int d^{4}\!x\left( \,\left( \frac{%
\delta \widetilde{\Sigma }}{\delta \Omega _{\mu }^{a}}+\frac{\delta
\widetilde{\Sigma }}{\delta \tau _{\mu }^{a}}\right) \frac{\delta \widetilde{%
\Sigma }}{\delta A_{\mu }^{a}}+\frac{\delta \widetilde{\Sigma }}{\delta
\Omega _{\mu }}\frac{\delta \widetilde{\Sigma }}{\delta A_{\mu }}+\frac{%
\delta \widetilde{\Sigma }}{\delta L^{a}}\frac{\delta \widetilde{\Sigma }}{%
\delta c^{a}}+\frac{\delta \widetilde{\Sigma }}{\delta L}\frac{\delta
\widetilde{\Sigma }}{\delta c}\right.  \nonumber \\
&&+b^{a}\frac{\delta \widetilde{\Sigma }}{\delta \bar{c}^{a}}+b\frac{\delta
\widetilde{\Sigma }}{\delta c}+\omega _{i}^{a}\frac{\delta \widetilde{\Sigma
}}{\delta \phi _{i}^{a}}+\bar{\phi}_{i}^{a}\frac{\delta \widetilde{\Sigma }}{%
\delta \bar{\omega}_{i}^{a}}+N_{\mu i}^{a}\frac{\delta \widetilde{\Sigma }}{%
\delta V_{\mu i}^{a}}-M_{\mu i}^{a}\frac{\delta \widetilde{\Sigma }}{\delta
U_{\mu i}^{a}}  \nonumber \\
&&\left. -(\Omega _{\mu }^{a}-\tau _{\mu }^{a})\frac{\delta \widetilde{%
\Sigma }}{\delta \xi _{\mu }^{a}}+\lambda _{i}^{a}\frac{\delta \widetilde{%
\Sigma }}{\delta \eta _{i}^{a}}+\rho _{i}^{a}\frac{\delta \widetilde{\Sigma }%
}{\delta \vartheta _{i}^{a}}+J\frac{\delta \widetilde{\Sigma }}{\delta
\Lambda }\right) \;.  \label{a7}
\end{eqnarray}
All other Ward identities, }eqs.$\left( \text{\ref{w}} -
\text{\ref{u1}}\right) $, turn
out to hold, remaining unmodified by the introduction of the sources $%
(J,\Lambda )$ . Moreover, there is an additional Ward identity{\
\begin{equation}
\mathcal{U}(\widetilde{\Sigma })=\int d^{4}\!x\,\left( \frac{\delta
\widetilde{\Sigma }}{\delta \Lambda }+\bar{c}^{a}\frac{\delta \widetilde{%
\Sigma }}{\delta b^{a}}\right) =0\;,  \label{a8}
\end{equation}
expressing the fact that the integrated operator }$\int d^{4}x\mathcal{O}%
_{A^{2}}$ is $BRST\;$invariant on-shell. \newline
\newline
By repeating the same analysis as before, for the most general local
counterterm we obtain
\begin{equation}
\widetilde{\Sigma }_{\mathrm{CT}}=a_{0}\,S_{\mathrm{YM}}+\mathcal{B}_{%
\widetilde{\Sigma }}\widetilde{\Delta }^{(-1)}\;,  \label{a9}
\end{equation}
with
\begin{eqnarray}
\widetilde{\Delta }^{(-1)} &=&\int d^{4}\!x\,\left( a_{1}\,[\Omega _{\mu
}^{a}A_{\mu }^{a}+\xi _{\mu }^{a}g\varepsilon ^{ab}A_{\mu }^{b}c]-a_{2}\,\xi
_{\mu }^{a}D_{\mu }^{ab}c^{b}+a_{3}\,L^{a}c^{a}-a_{4}\,\bar{c}^{a}D_{\mu
}^{ab}A_{\mu }^{b}\right.  \nonumber \\
&&+a_{5}\,Lc+a_{6}\,\chi \,U_{\mu i}^{a}V_{\mu
i}^{a}+(a_{1}+a_{2}+a_{3})[U_{\mu i}^{a}D_{\mu }^{ab}\phi _{i}^{b}-V_{\mu
i}^{a}D_{\mu }^{ab}\bar{\omega}_{i}^{b}]-(a_{3}+a_{4})\,\bar{\omega}_{i}^{a}%
\mathcal{M}^{ab}\phi _{i}^{b}  \nonumber \\
&&+\frac{\alpha }{2}a_{7}\left( \bar{c}^{a}b^{a}-g\varepsilon ^{ab}\bar{c}%
^{a}\bar{c}^{b}c\right) +\frac{\alpha }{2}\left( a_{7}-2a_{3}\right) g^{2}%
\bar{\omega}_{i}^{a}\phi _{i}^{a}\left( \bar{\phi}_{j}^{b}\phi _{j}^{b}-\bar{%
\omega}_{j}^{b}\omega _{j}^{b}\right) -\alpha \left( a_{7}-a_{3}\right) g^{2}%
\bar{\omega}_{i}^{a}\phi _{i}^{a}\overline{c}^{b}c^{b}  \nonumber \\
&&\left. +\Lambda \left( \frac{1}{2}(a_{3}-a_{4})\,A_{\mu
}^{a}A_{\mu }^{a}\;+\alpha \left( a_{7}+a_{3}\right)
\overline{c}^{a}c^{a}-\alpha a_{7}\left( \bar{\phi}_{j}^{b}\phi
_{j}^{b}-\bar{\omega}_{j}^{b}\omega _{j}^{b}\right)
+a_{8}\,\frac{\varrho }{2}J\;\right) \right) \;. \label{cct}
\end{eqnarray}
In particular, for the renormalization of the sources $J$ and
$\Lambda $, and of the parameter $\varrho $ one finds
\begin{eqnarray}
J_{0} &=&Z_{J}\,J=Z_{g}^{2}Z_{\bar{c}}^{-1}\,J\;,  \nonumber \\
\Lambda _{0} &=&Z_{\Lambda }\,\Lambda =Z_{g}Z_{\bar{c}}^{-1/2}\,\Lambda \;,
\nonumber \\
\varrho_{0} &=&Z_{\varrho }\,\varrho  =\left( 1+\epsilon
(2a_{0}-2a_{3}+2a_{4}+a_{8})\,\right) \varrho \;. \label{a11}
\end{eqnarray}
We see thus that the renormalization of the source $J$, and thus
of the composite operator coupled to it, \textit{i.e.}
$\mathcal{O}_{A^{2}}$, can be expressed in terms of the
renormalization factors of the gauge coupling constant $g$ and of
the diagonal antighost $\bar{c}$, meaning that the anomalous
dimension of $\mathcal{O}_{A^{2}}$ is not an independent parameter
of the theory. Again, this result is in complete analogy with the
case of the Landau gauge, where the corresponding operator $A_{\mu
}^{A}A_{\mu }^{A}$
is multiplicatively renormalizable\footnote{%
Here too, the anomalous dimension of $A_{\mu }^{A}A_{\mu }^{A}$ is not an
independent parameter of the theory \cite{Dudal:2005na}.} \cite{Dudal:2005na}
in the presence of Zwanziger's horizon function $\left( \ref{i3}\right) $.

\section{Conclusion}

In this work we have pursued the study of the maximal Abelian gauge by
taking into account the restriction of the domain of integration in the
Faddeev-Popov quantization formula to the Gribov region $\mathcal{C}_{0}$.
Such a restriction is needed due to the existence of Gribov copies. The
Gribov approximation, previously introduced in \cite{Capri:2005tj}, has been
improved through the introduction of the nonlocal horizon function, eq.$%
\left( \ref{i8}\right) $. As in the case of the Landau gauge \cite
{Zwanziger:1989mf,Zwanziger:1992qr}, the horizon term of the maximal Abelian
gauge can be cast in local form with the help of additional auxiliary
fields. The resulting local action $\Sigma $, eq.$\left( \text{\ref{sigma}}%
\right) $, turns out to be multiplicatively renormalizable to all orders.
This is the main result of the present article. It could open new
perspectives, motivating further analytic studies of the maximal Abelian
gauge which might provide a better comparison with lattice numerical results
\cite{Amemiya:1998jz,Bornyakov:2003ee}. A partial list of the topics worth
to be analysed can be summarized as follows:

\begin{itemize}
\item  The possibility of having at our disposal a local and
renormalizable action incorporating the effects of the Gribov
horizon enables us to work out the higher order quantum
corrections affecting both the diagonal and off-diagonal gluon
propagator. We expect that the incorporation of these quantum
corrections will provide a more direct and reliable comparison
with the lattice data. As an example, we mention the possibility
of investigating by analytical methods the behavior of the
longitudinal off-diagonal component of the gluon propagator, which
we were unable to discuss within Gribov\'{}s quadratic
approximation \cite{Capri:2005tj}. Lattice results
\cite{Bornyakov:2003ee} have pointed out that this component is
nonvanishing. We remark, however, that, from the analytic point of
view, it vanishes at the tree level. Nevertheless, it might arise
at the quantum level due to the contribution of the diagonal and
off-diagonal transverse components of the gluon propagator in the
loop integrals of higher order Feynman diagrams. As these
transverse components carry nonperturbative information, embodied
in the Gribov parameter $\gamma $ and in the dynamical gluon mass
$m$, a nonvanishing longitudinal component might show up at the
quantum level. Certainly, the renormalizability of the starting action $%
\Sigma $ is a crucial ingredient here.

\item  A second issue to be investigated is the infrared behavior of the
off-diagonal ghost propagator $\left\langle \bar{c}^{a}(k)c^{b}(-k)\right%
\rangle $, which can be decomposed into symmetric and
antisymmetric part with respect to the off-diagonal indices
$a,b=1,2$. As already mentioned, the symmetric part of the ghost
propagator, \textit{i.e. }$\sum_{a,b}\delta ^{ab}\left\langle
\bar{c}^{a}(k)c^{b}(-k)\right\rangle $, has been found to be
enhanced in the infrared region, in the Gribov approximation
\cite{Capri:2005tj}, as expressed by eq.$\left(
\ref{offghg}\right) $. A more detailed study of the behavior of
the symmetric component of the ghost propagator would provide an
important check in order to establish whether this infrared
enhancement will be left unmodified by the higher order
corrections. \newline
\newline
Also, the study of the antisymmetric part of the ghost propagator, \textit{%
i.e. }$\varepsilon ^{ab}\left\langle \bar{c}^{a}(k)c^{b}(-k)\right\rangle $,
would provide information about the phenomenon of the ghost condensation in
the presence of the Gribov horizon. The possibility of the formation of a
ghost condensate $\left\langle \varepsilon ^{ab}\bar{c}^{a}(x)c^{b}(x)\right%
\rangle $ in the maximal Abelian gauge, proposed in \cite{Schaden:1999ew},
has been investigated by several authors \cite
{Kondo:2001nq,Lemes:2002ey,Dudal:2002xe}. This phenomenon has been shown to
occur also in the Landau gauge \cite{Lemes:2002rc,Dudal:2003dp,Capri:2005vw}%
, where a few lattice results are available \cite
{Cucchieri:2005yr,Furui:2006rx}. Here, the ghost condensation has been
established by constructing the effective potential for the operator $%
\varepsilon ^{ab}\bar{c}^{a}(x)c^{b}(x)$. The existence of a nontrivial
minimum for the effective potential resulted in a nonvanishing condensate $%
\left\langle \varepsilon ^{ab}\bar{c}^{a}c^{b}\right\rangle $ \cite
{Lemes:2002rc,Dudal:2003dp,Capri:2005vw}. It could be interesting to see how
the phenomenon of the ghost condensation would be modified by the presence
of the Gribov horizon, a feature which can be now faced. This might result
in a better comparison with the available lattice results\footnote{%
We are indebted to A. Cucchieri and T. Mendes for providing us their
preliminary results about the numerical studies on the lattice of the ghost
condensation in the maximal Abelian gauge.} on the ghost condensation, which
are done by considering always gauge configurations which lie within the
Gribov region. The construction of the effective potential for the ghost
operator $\varepsilon ^{ab}\bar{c}^{a}(x)c^{b}(x)$ in the presence of the
Gribov horizon is under investigation. Both Landau and maximal Abelian gauge
will be covered.

\item  The multiplicative renormalizability of the local composite operator $%
\mathcal{O}_{A^{2}}$, eq.$\left( \ref{oa2}\right) $,$\;$ opens the
possibility to perform a study of the dimension two gluon condensate $%
\left\langle \mathcal{O}_{A^{2}}\right\rangle $ when the restriction to the
Gribov region $\mathcal{C}_{0}$ is taken into account. This might improve
our understanding of the dynamical mass generation for off-diagonal gluons,
a feature relevant for the dual superconductivity picture of color
confinement.

\item  Finally, we would like to call attention to the nonrenormalization
theorem
\begin{equation}
Z_{A}Z_{g}^{2}=1\;,  \label{nrt}
\end{equation}
which still holds in the presence of the horizon. This result,
stemming from the local $U(1)$ Ward identity $\left(
\ref{u1}\right) $, could motivate further studies of the maximal
Abelian gauge from both Schwinger-Dyson approach and lattice
numerical simulations. Concerning the potential applications to
the Schwinger-Dyson equations, we remark that eq.$\left( \ref
{nrt}\right) $ could be employed in order to obtain a useful
truncation scheme. Also, it would be interesting to look at
relation $\left( \ref {nrt}\right) $ from the lattice point of
view, as it could allow to study the infrared behavior of the
running coupling constant by analysing the form factor of the pure
diagonal gluon propagator.
\end{itemize}

\section*{Acknowledgments}

The Conselho Nacional de Desenvolvimento Cient\'{i}fico e Tecnol\'{o}gico
(CNPq-Brazil), the Faperj, Funda{\c{c}}{\~{a}}o de Amparo {\`{a}} Pesquisa
do Estado do Rio de Janeiro, the SR2-UERJ and the Coordena{\c{c}}{\~{a}}o de
Aperfei{\c{c}}oamento de Pessoal de N{\'{i}}vel Superior (CAPES) are
gratefully acknowledged for financial support.

\end{document}